\documentclass[twocolumn,showpacs,prb,superscriptaddress,a4paper,floatfix]{revtex4}
\usepackage{amssymb}
\usepackage{graphicx}
\usepackage{amsmath}
\usepackage{graphicx}
\usepackage{dcolumn}
\usepackage{bm}
\usepackage{color}

\usepackage{multirow}             %table
\usepackage{setspace}             % table
\usepackage{siunitx}
\usepackage{booktabs}
\usepackage{array}
\newcolumntype{C}[1]{>{\centering\arraybackslash}p{#1}}
\begin{document}

\title{Fingerprints of disorder in graphene}
\author{Pei-Liang Zhao}
\affiliation{Department of Applied Physics, Zernike Institute for Advanced Materials,
University of Groningen, Nijenborgh 4, NL-9747AG Groningen, The Netherlands}
\author{Shengjun Yuan}
\email{s.yuan@science.ru.nl}
\affiliation{Radboud University, Institute for Molecules and Materials,
Heijendaalseweg 135, 6525AJ Nijmegen, The Netherlands}
\author{Mikhail I. Katsnelson}
\affiliation{Radboud University, Institute for Molecules and Materials,
Heijendaalseweg 135, 6525AJ Nijmegen, The Netherlands}
\author{Hans De Raedt}
%\email{h.a.de.raedt@rug.nl}
\affiliation{Department of Applied Physics, Zernike Institute for Advanced Materials,
University of Groningen, Nijenborgh 4, NL-9747AG Groningen, The Netherlands}
\date{\today }

\begin{abstract}
We present a systematic study of the electronic, transport and optical
properties of disordered graphene including the next-nearest-neighbor
hopping. We show that this hopping has a non-negligible effect on resonant
scattering but is of minor importance for long-range disorder such as
charged impurities, random potentials or hoppings induced by strain
fluctuations. Different types of disorders can be recognized by their
fingerprints appearing in the profiles of dc conductivity, carrier mobility,
optical spectroscopy and Landau level spectrum. The minimum conductivity $%
4e^{2}/h$ found in the experiments is dominated by long-range disorder and
the value of $4e^{2}/\pi h$ is due to resonant scatterers only.
\end{abstract}

\maketitle

%\email{h.a.de.raedt@rug.nl}

%\email{h.a.de.raedt@rug.nl}

%\email{h.a.de.raedt@rug.nl}

\section{Introduction}

\begin{figure}[t]
\begin{center}
\mbox{
 \includegraphics[width=0.41\columnwidth]{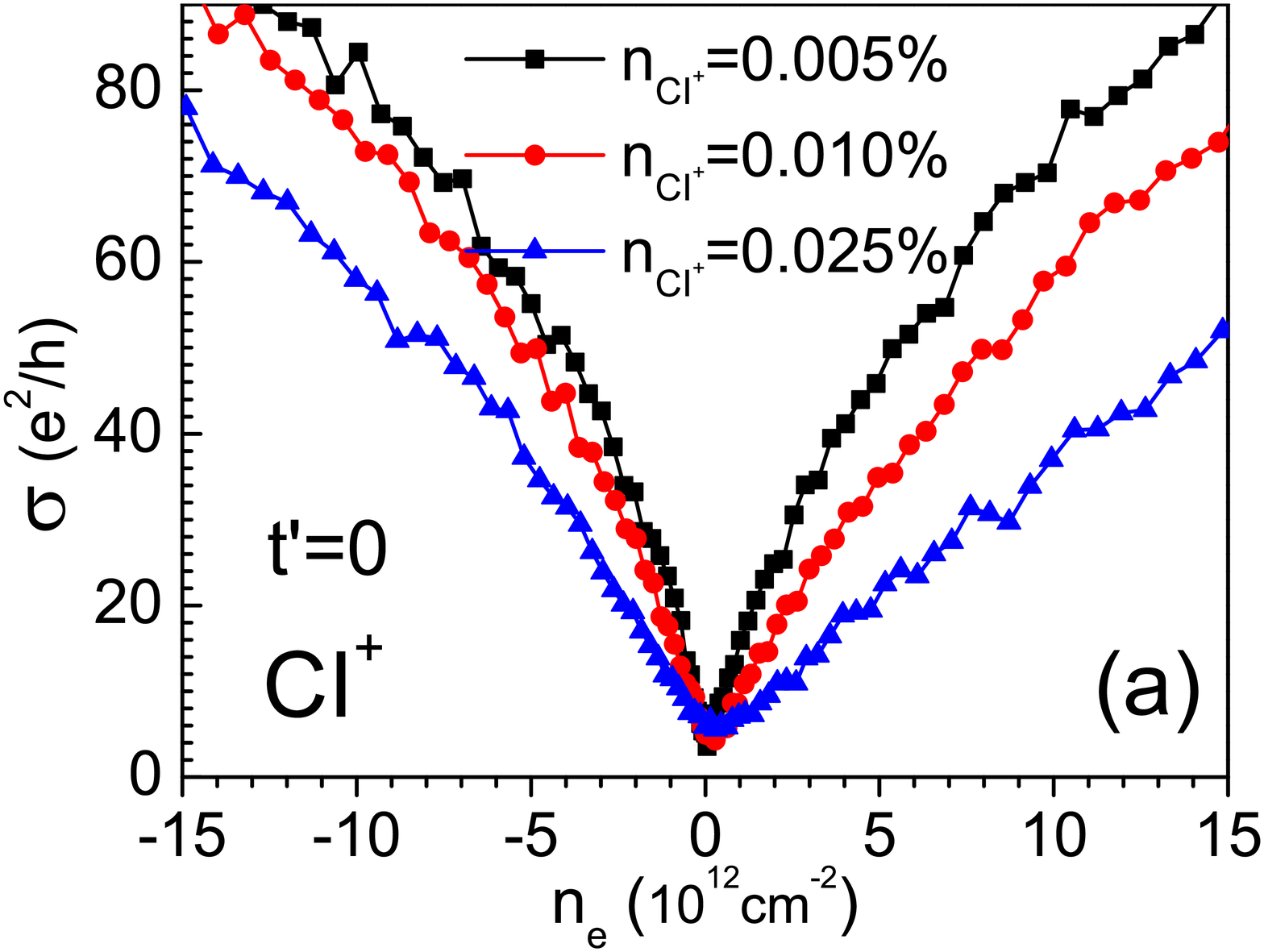}
 \includegraphics[width=0.41\columnwidth]{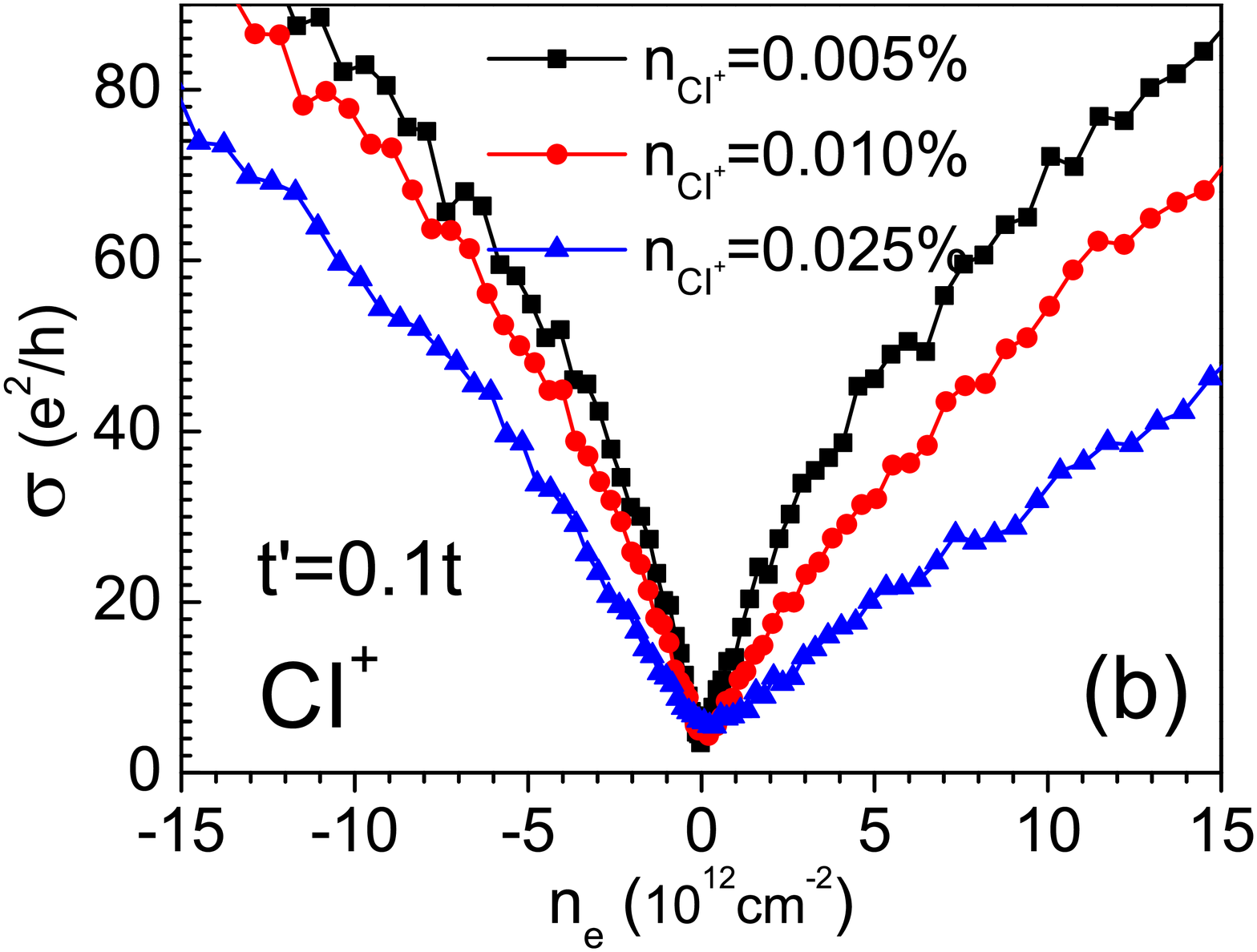}
  }
\mbox{
 \includegraphics[width=0.41\columnwidth]{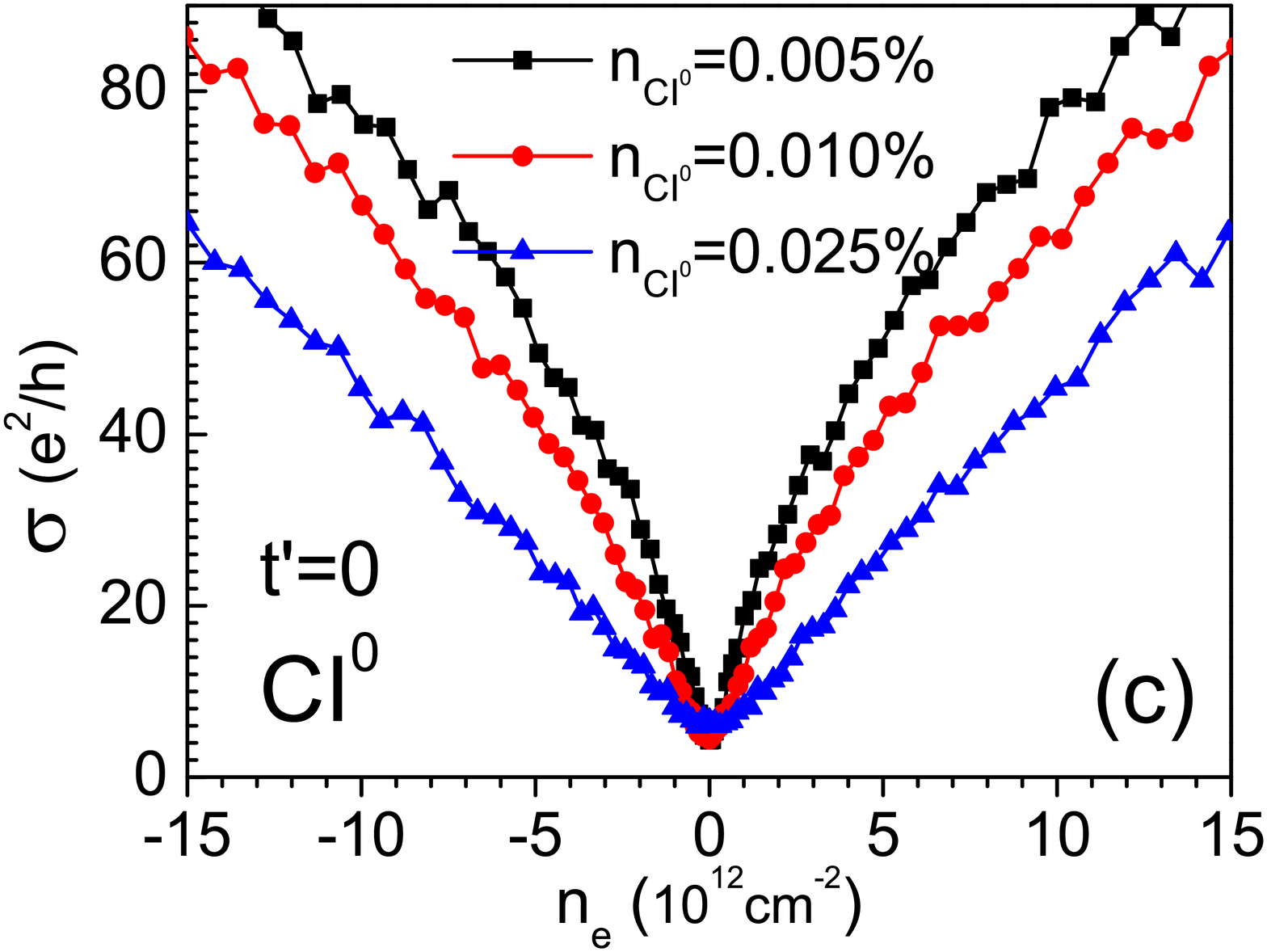}
 \includegraphics[width=0.41\columnwidth]{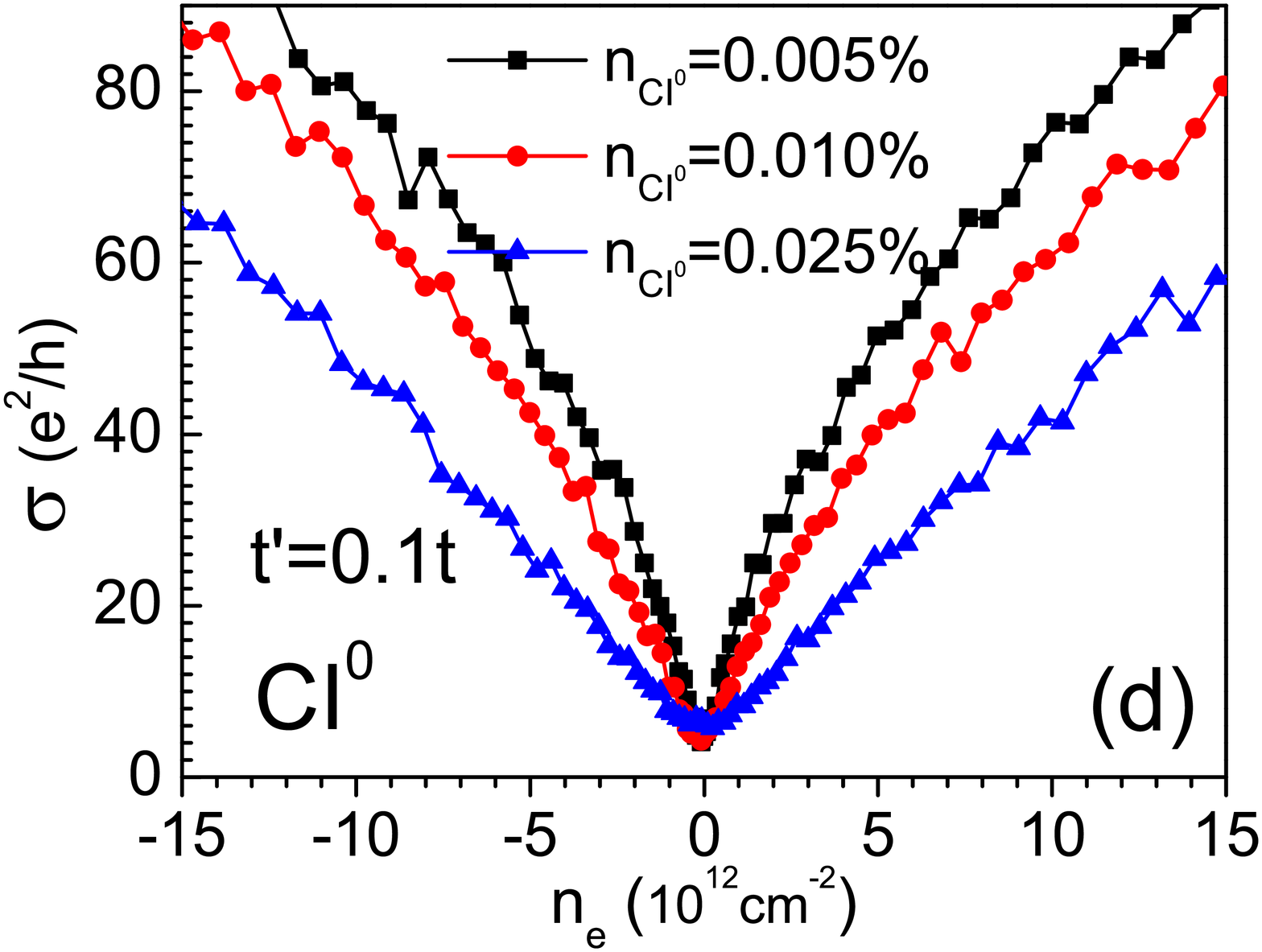}
  }
\mbox{
 \includegraphics[width=0.41\columnwidth]{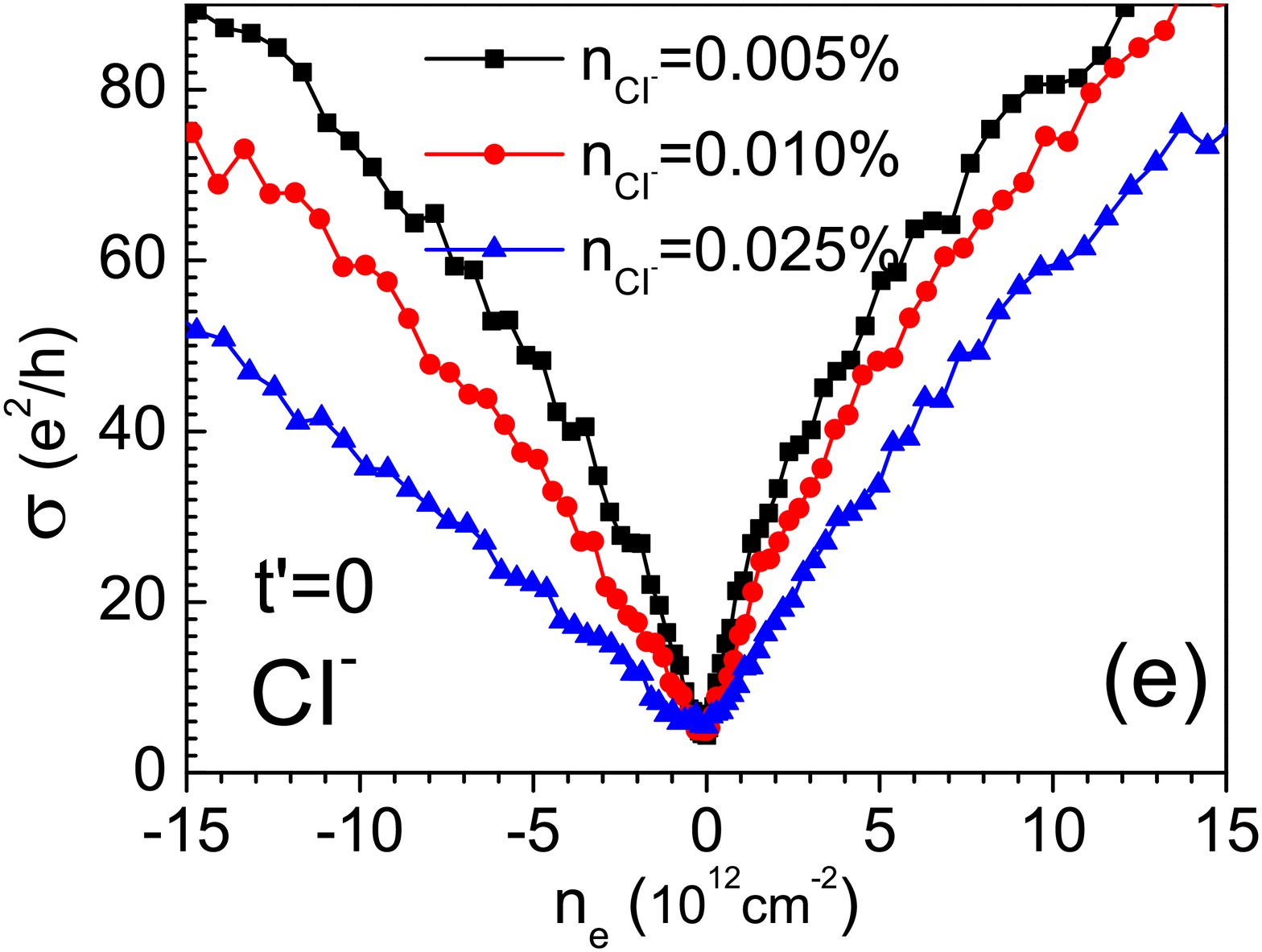}
 \includegraphics[width=0.41\columnwidth]{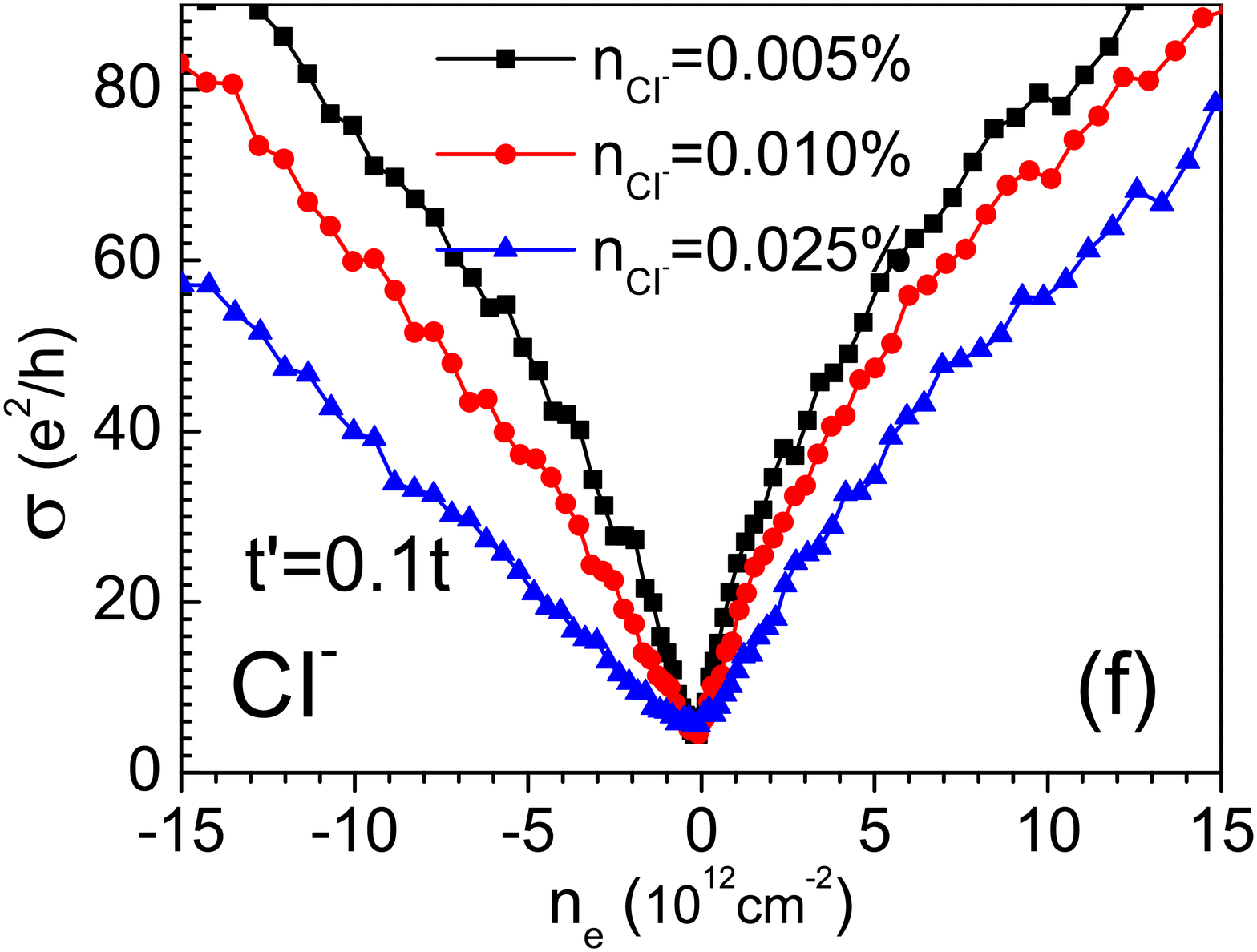}
  }
\mbox{
 \includegraphics[width=0.41\columnwidth]{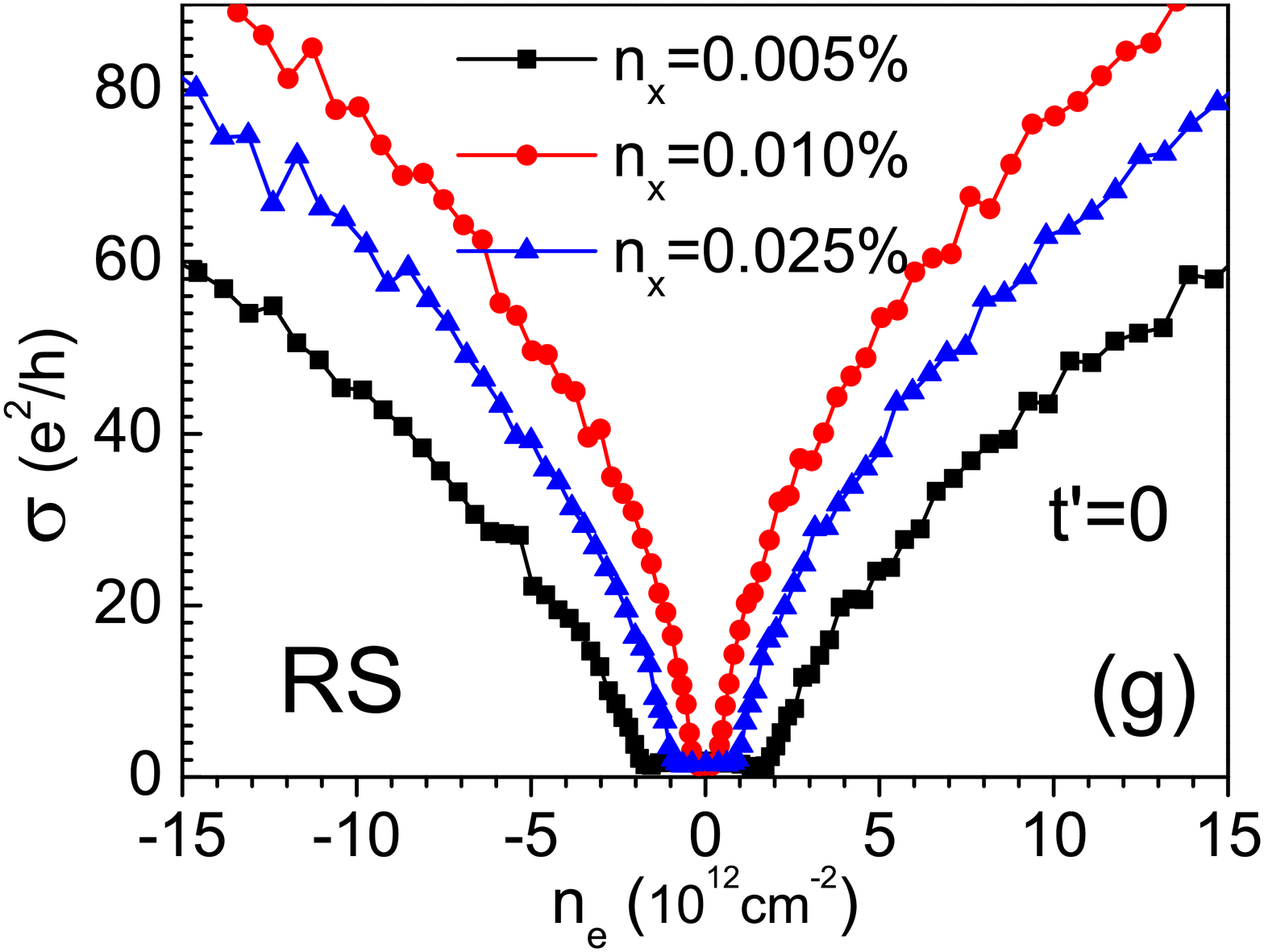}
 \includegraphics[width=0.41\columnwidth]{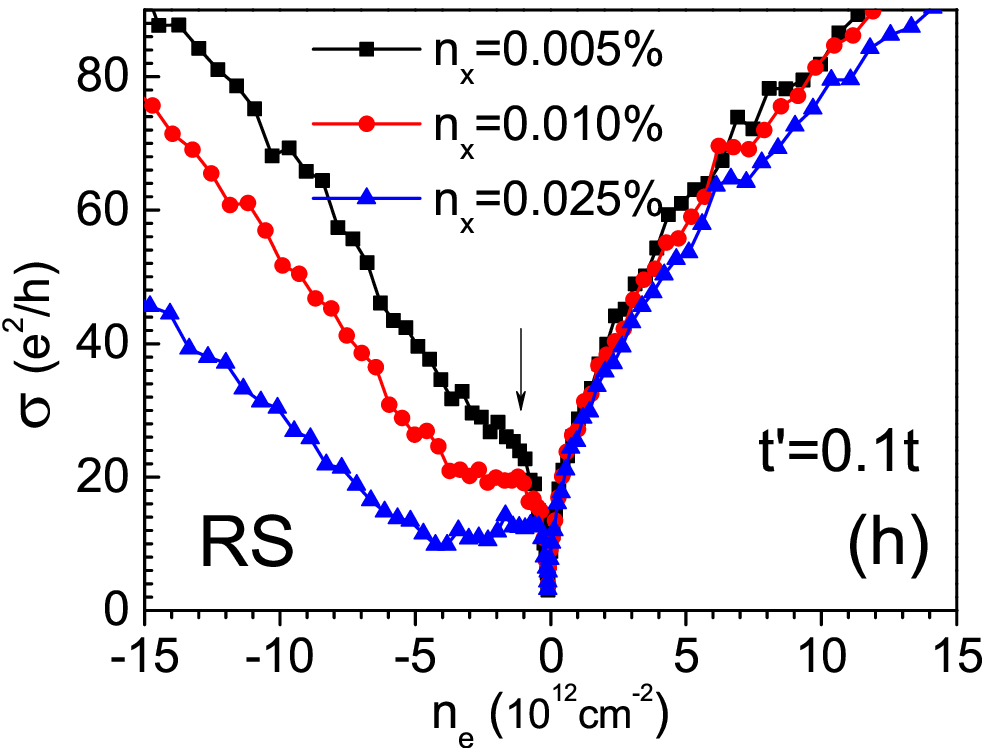}
  }
\mbox{
 \includegraphics[width=0.41\columnwidth]{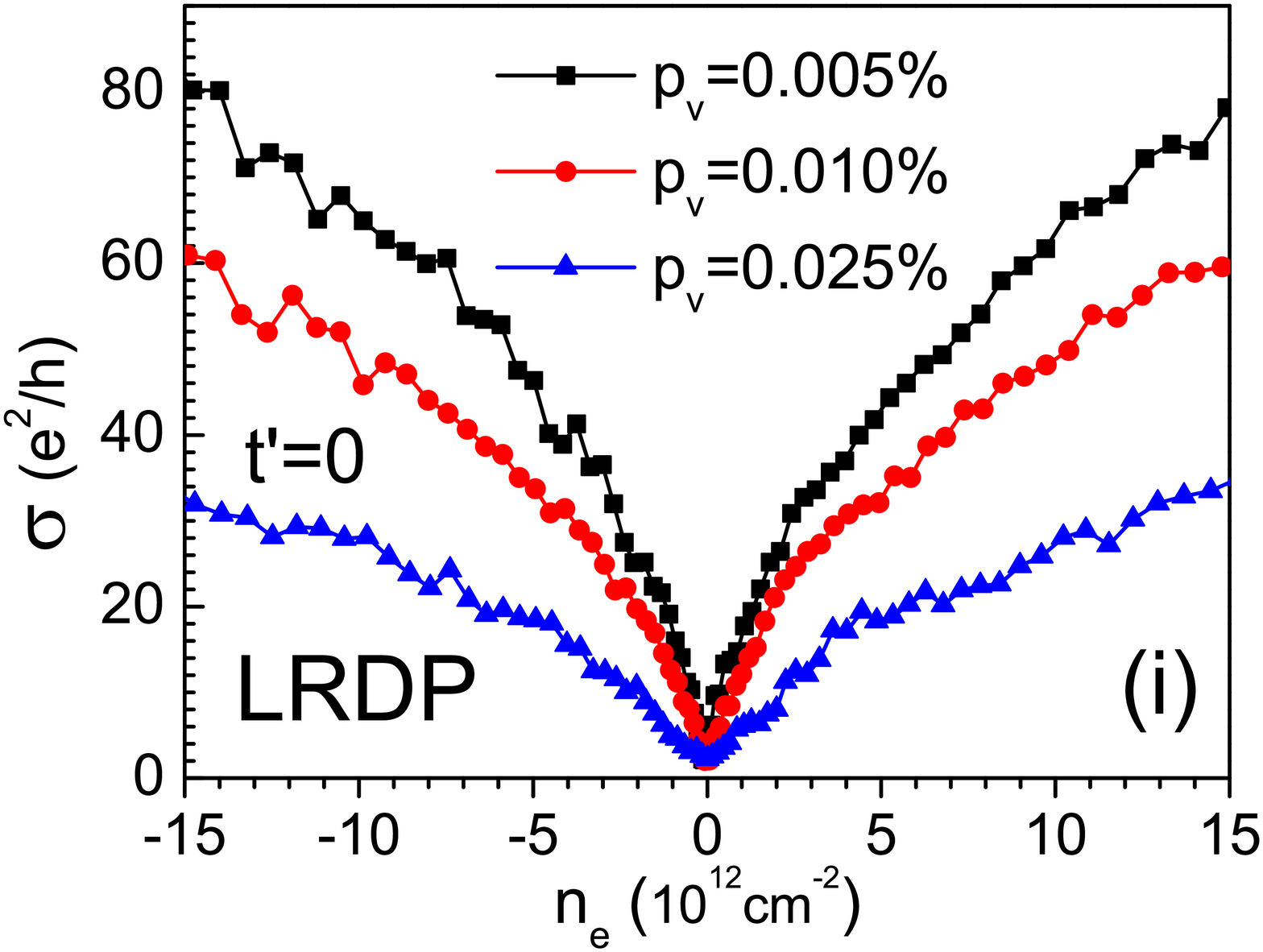}
 \includegraphics[width=0.41\columnwidth]{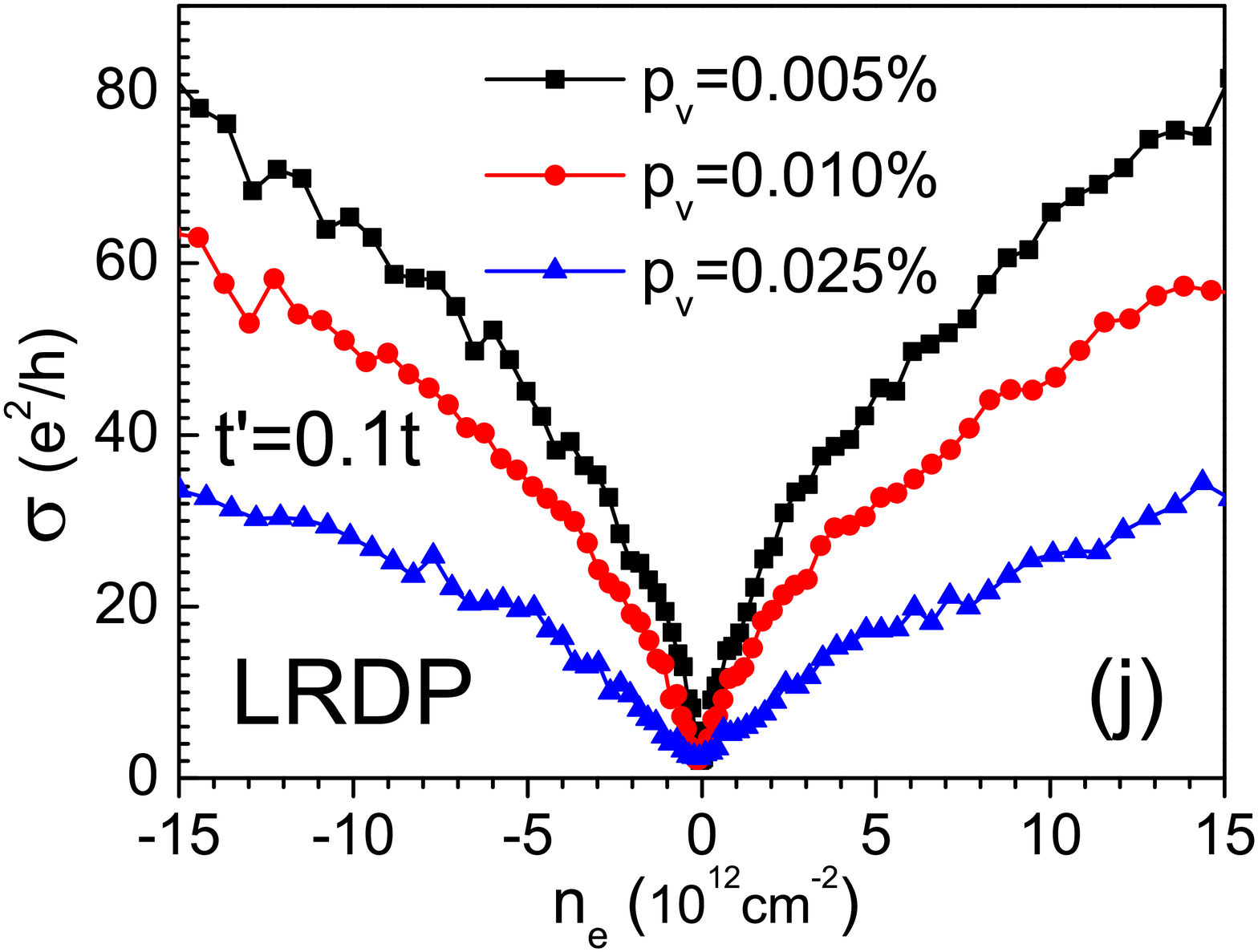}
  }
\mbox{
 \includegraphics[width=0.41\columnwidth]{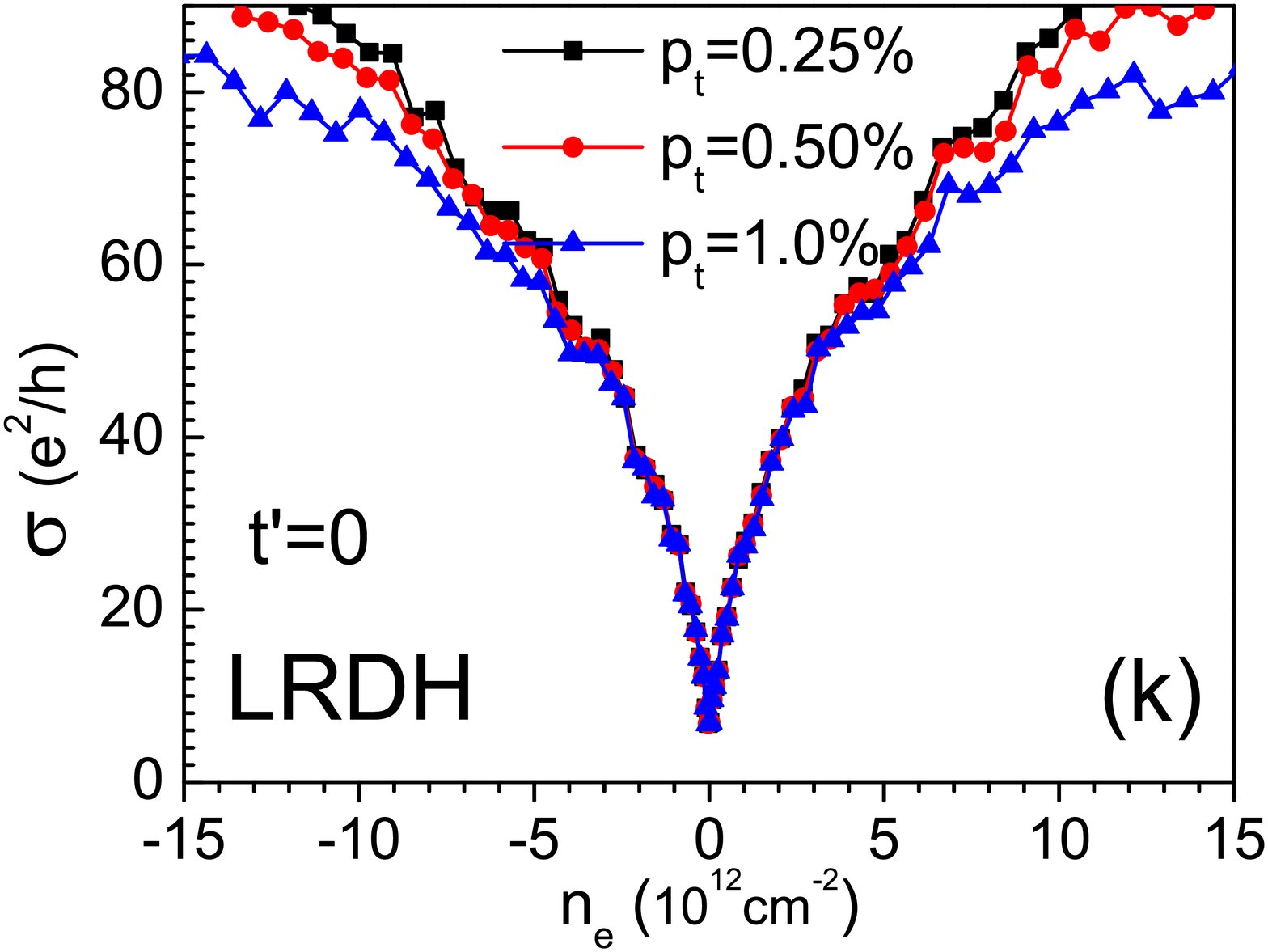}
 \includegraphics[width=0.41\columnwidth]{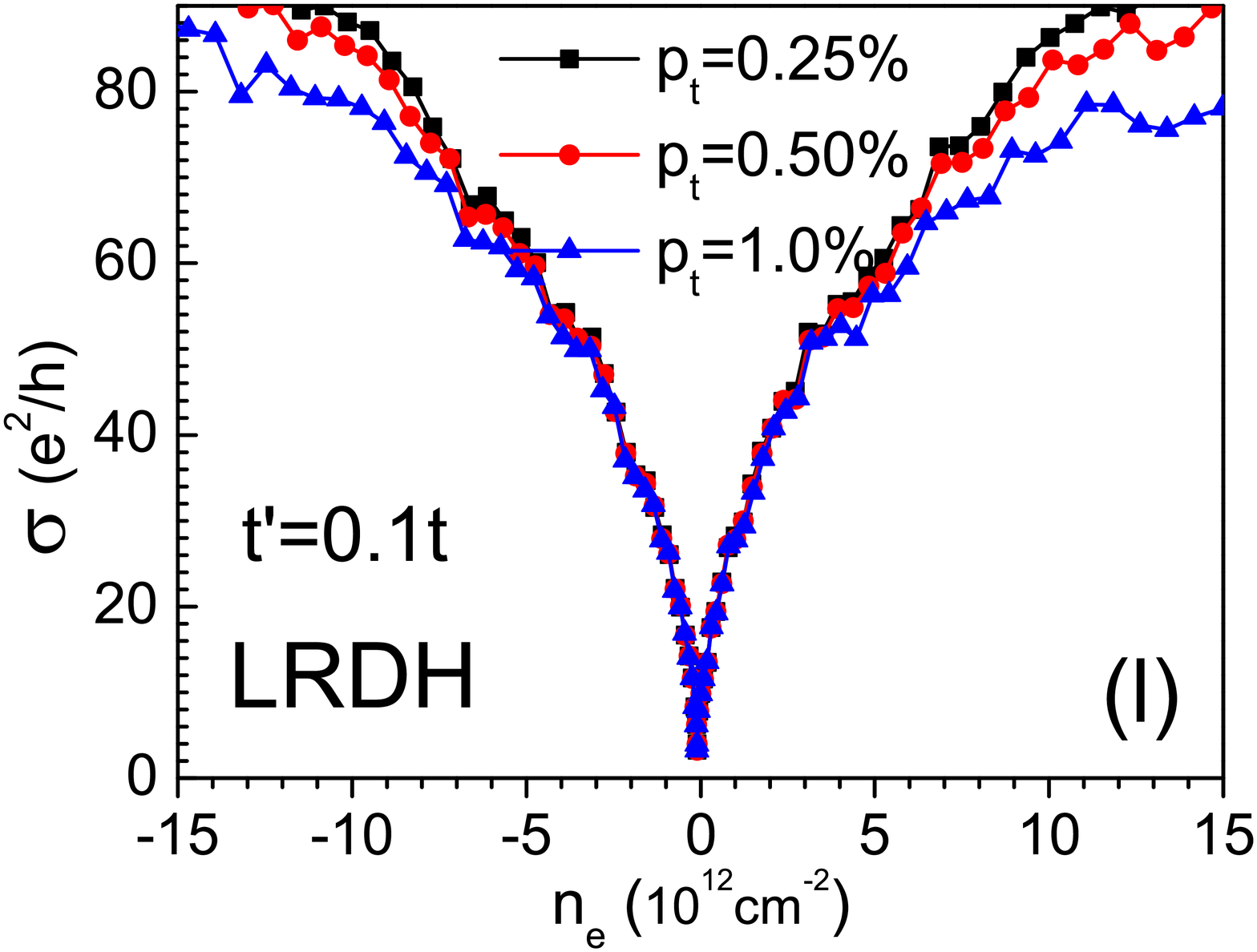}
  }
\end{center}
\caption{ (Color online) The dc conductivity as a function of carrier
density $n_{e}$ for disordered graphene. Left panels show the results
without the NNN hopping $t^{\prime }$, and right panels with $t^{\prime
}=0.1t$. For CI, we use $\protect\kappa =6$ of hexagonal-boron nitride as a
typical value of dielectric constant for graphene on a substrate. The use of
other $\protect\kappa $ for different substrate such as SiO$_{2}$ does not
change the results quantitatively. Here $0.01\%$ disorder corresponds to a
concentration of $3.82\times 10^{11}cm^{-2}$. }
\label{Fig:DC}
\end{figure}

\begin{figure}[h]
\begin{center}
\mbox{
 \includegraphics[width=0.48\columnwidth]{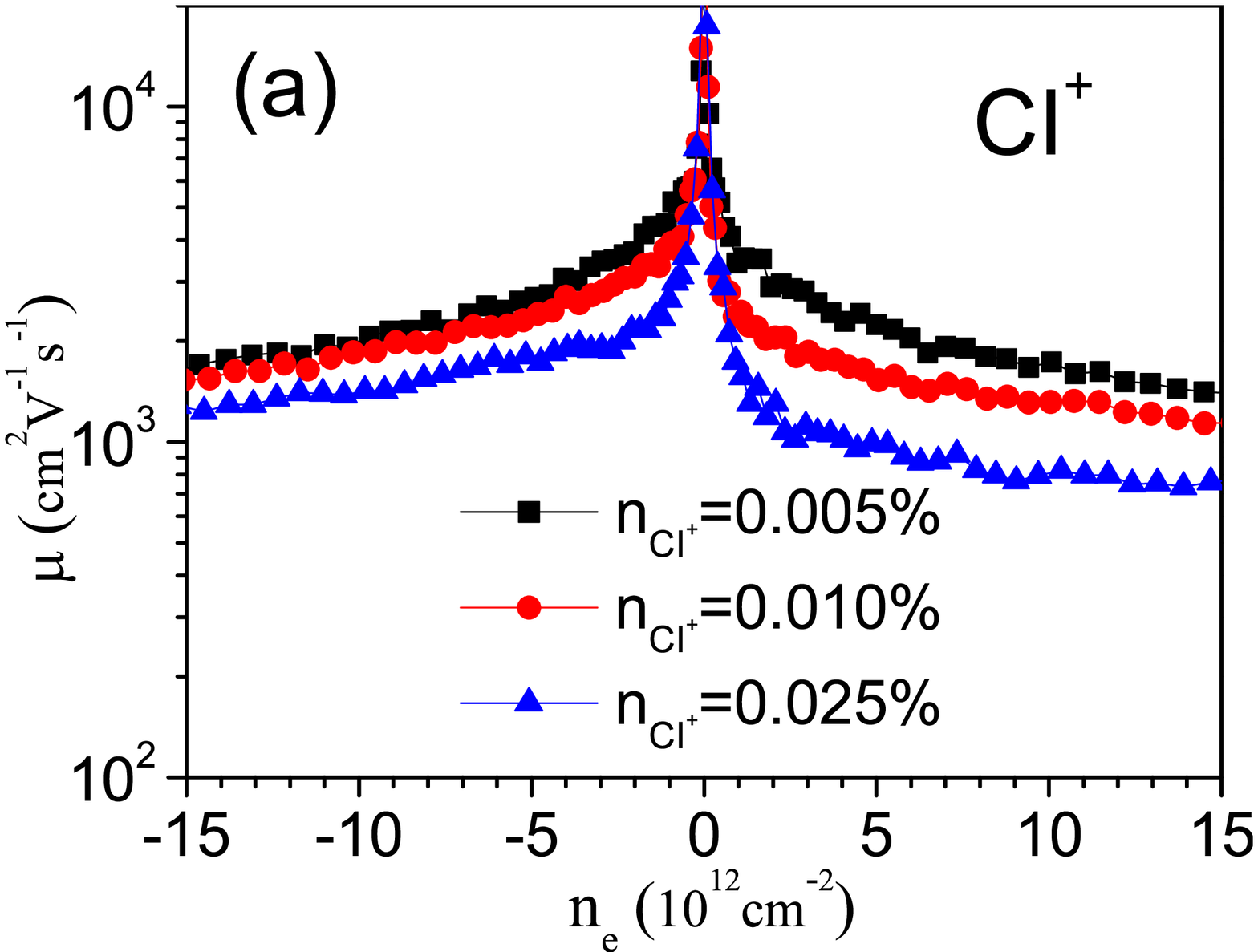}
 \includegraphics[width=0.48\columnwidth]{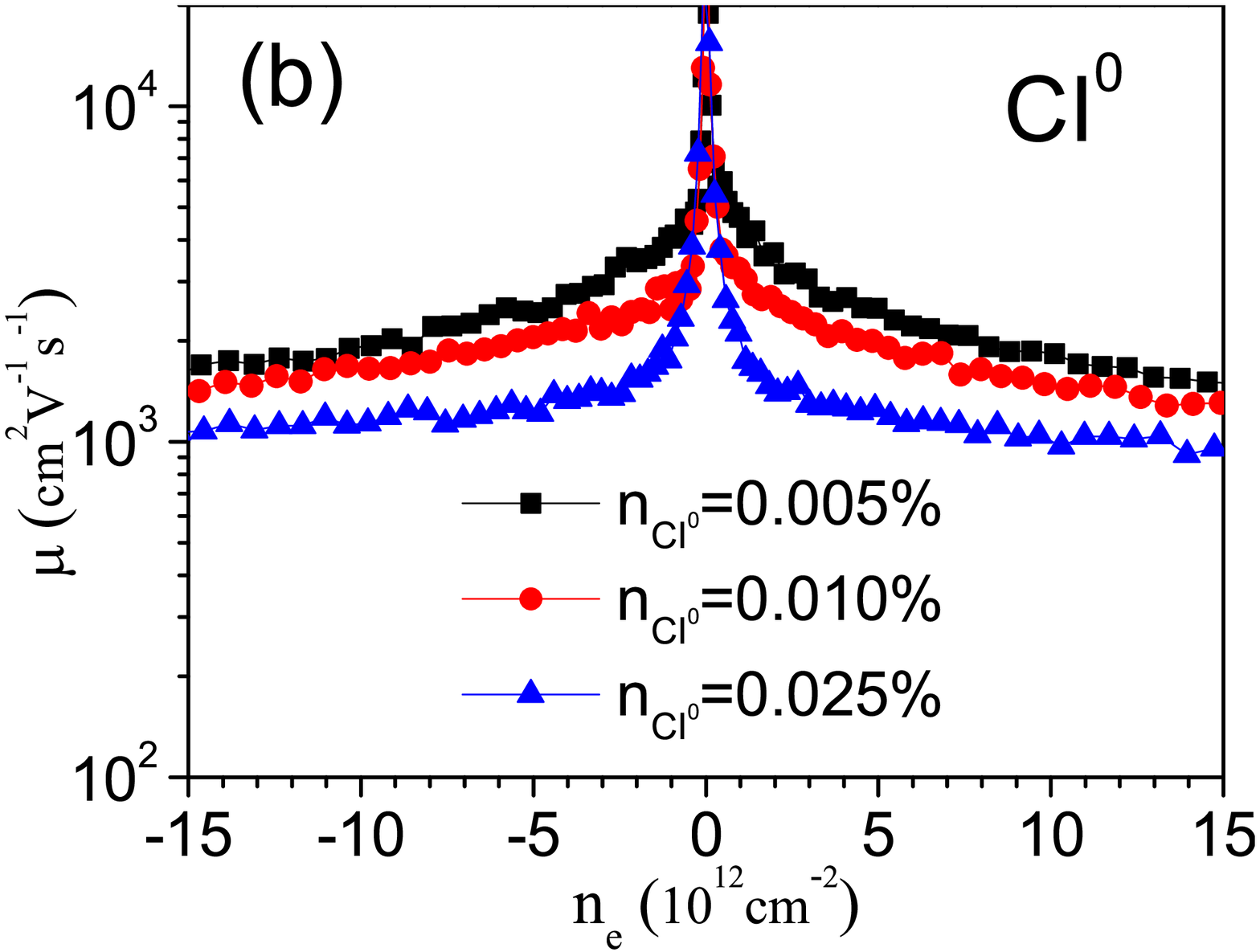}
  }
\mbox{
 \includegraphics[width=0.48\columnwidth]{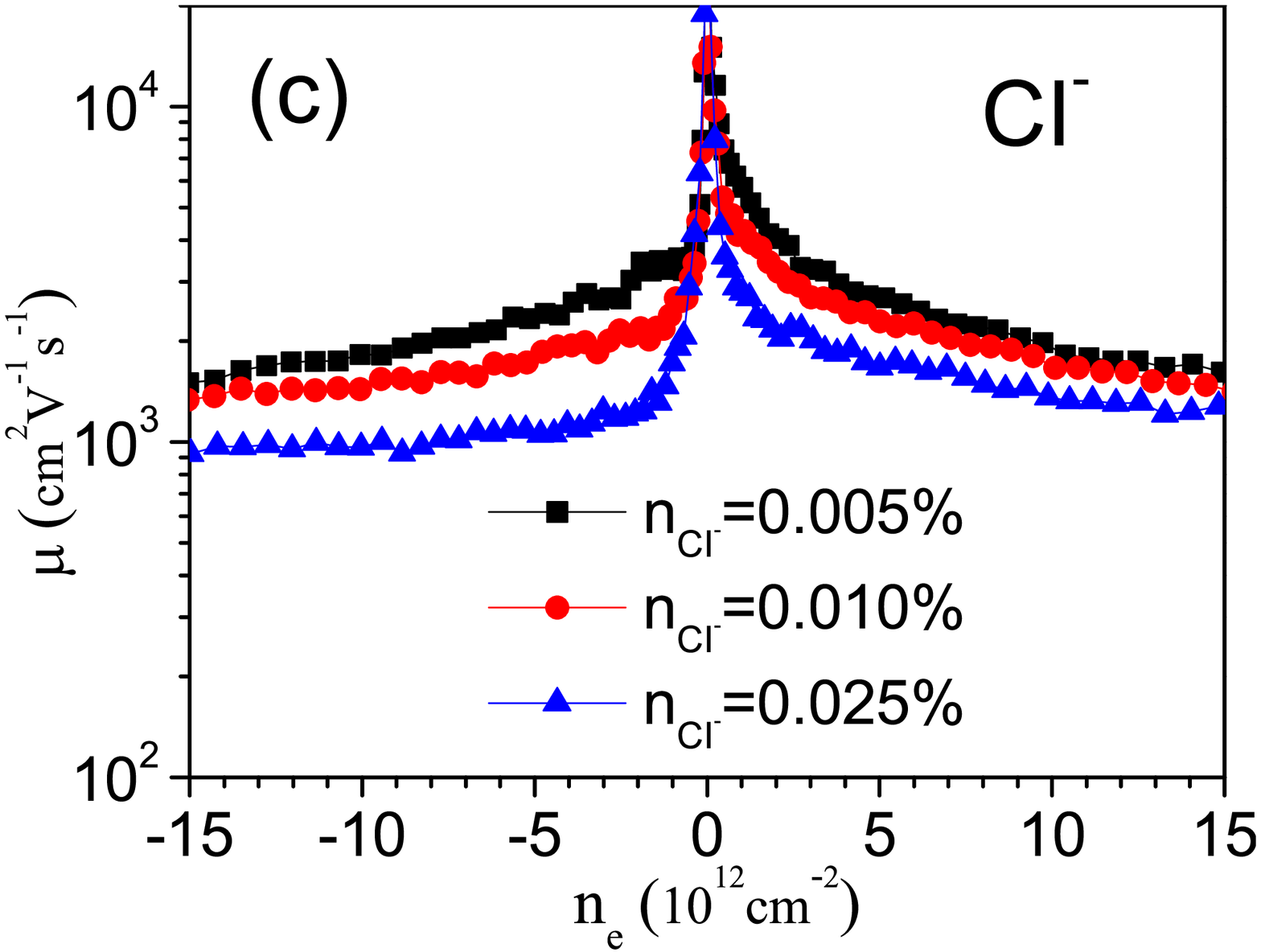}
 \includegraphics[width=0.48\columnwidth]{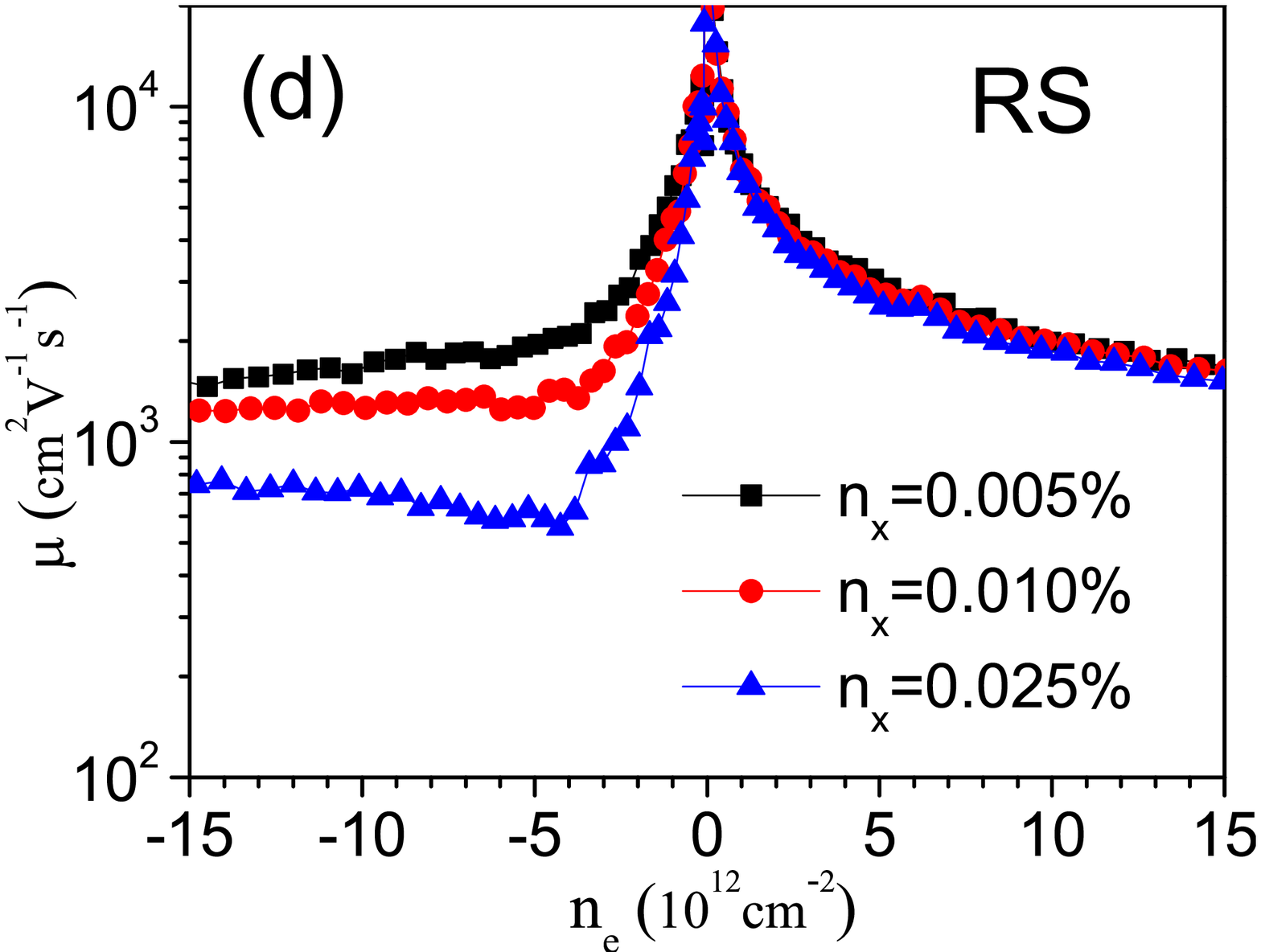}
  }
\mbox{
 \includegraphics[width=0.48\columnwidth]{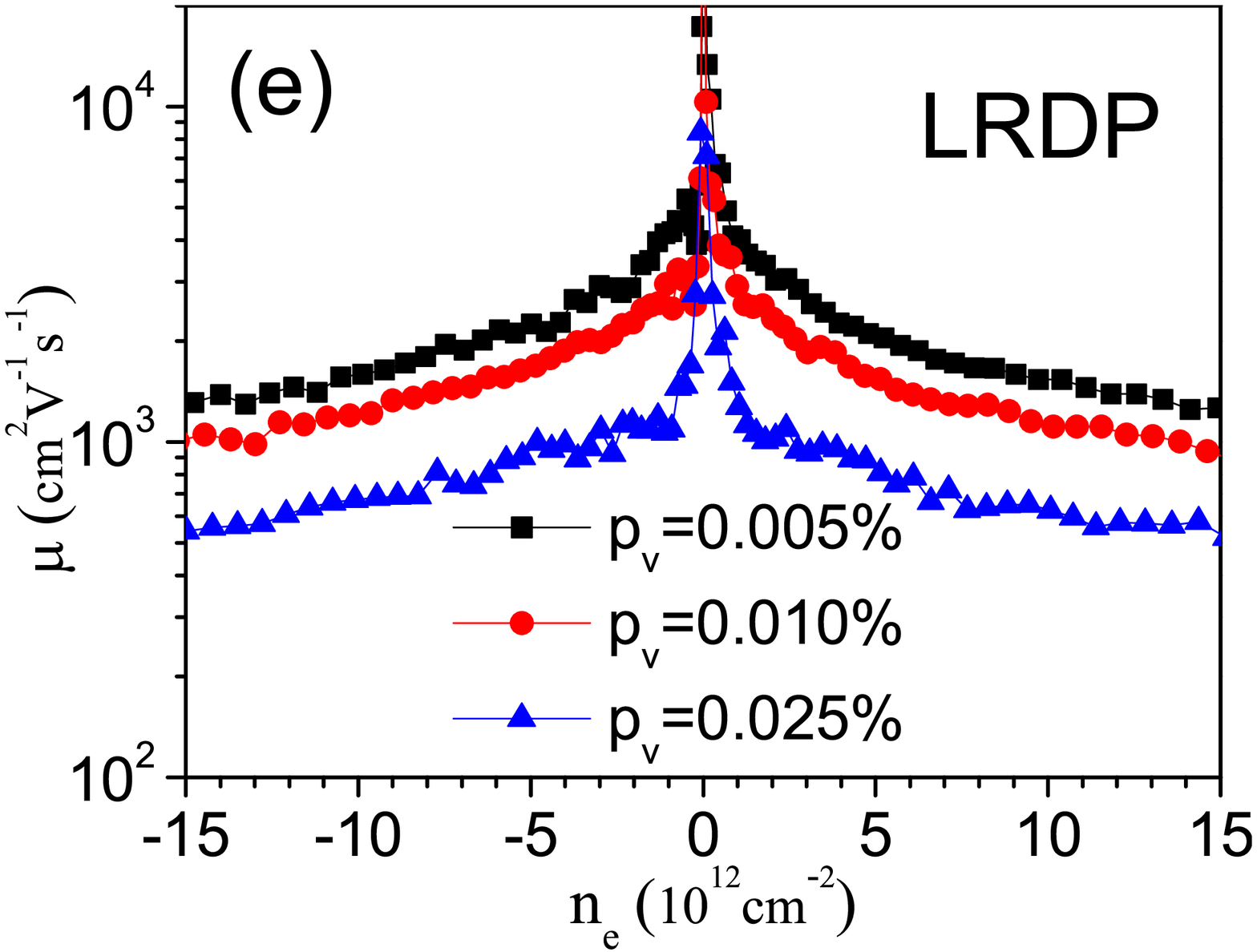}
 \includegraphics[width=0.48\columnwidth]{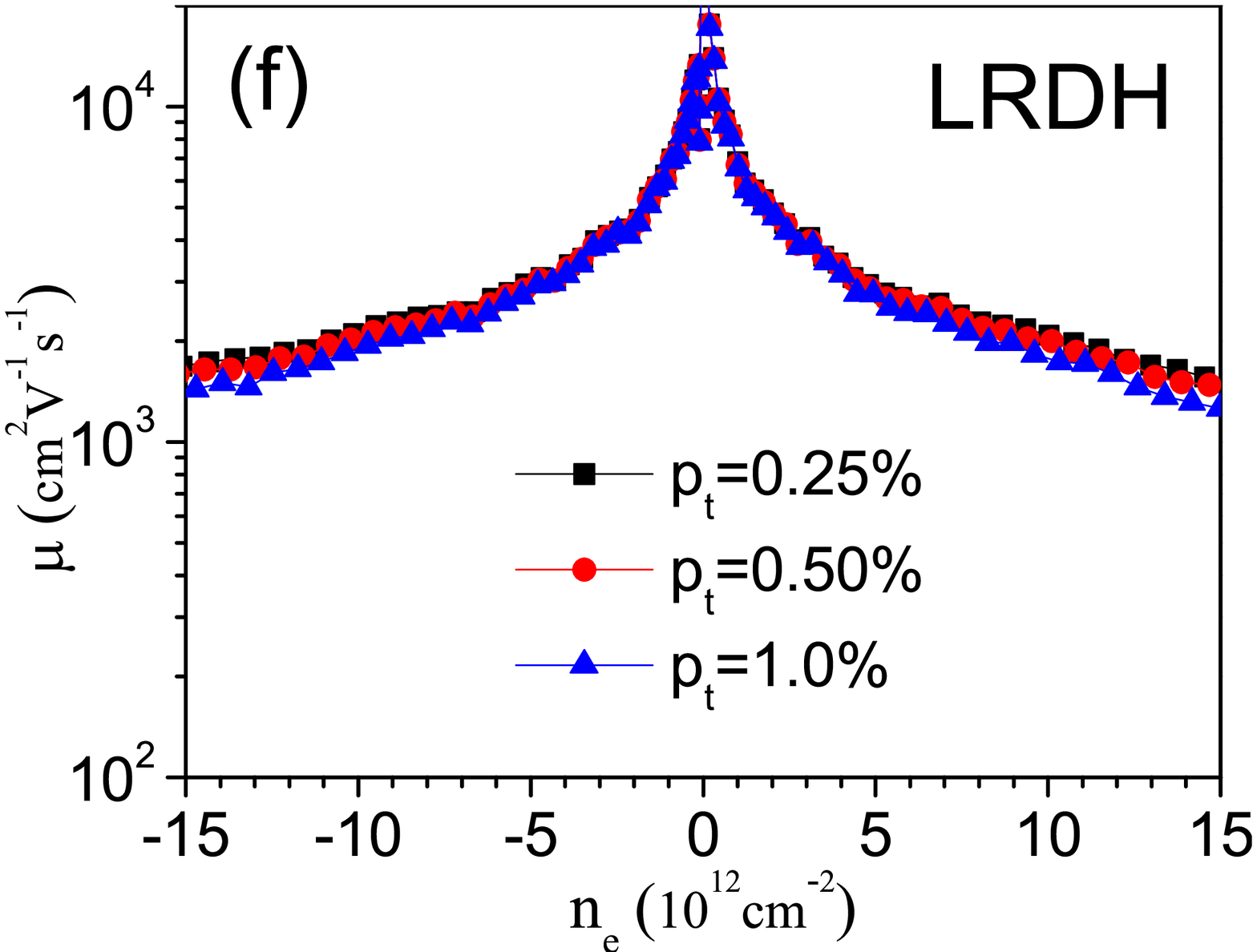}
  }
\end{center}
\caption{ (Color online) The carrier mobility as a function of carrier
density $n_{e}$ for disorder graphene with $t^{\prime }$.}
\label{Fig:mobility}
\end{figure}

The dominant source of disorder which limits the transport and optical
properties of graphene is still under debate. Different mechanisms have been
proposed and investigated intensively, including charged impurities, random
strain fluctuations and resonant scatterers (for reviews see Refs.~%
\onlinecite{Peres2010RMP,KatsnelsonBook}). Early on, charged impurities (CI)
have been recognized as the dominate disorders due to graphene's unusual
linear carrier-density-dependent conductivity. However, this mechanism does
not explain the experimental observations that the transport properties of
certain samples are not sensitive to the substrate screening \cite%
{Ponomarenko2009,Couto2011}. On the other hand,
strain fluctuations (SF) induced e.g. ripples
can be alternative scattering mechanism ~\cite%
{ripple2008}; they can be also responsible for charge inhomogeneities, that is,
electron-hole puddles~\cite{Gibertini2010,Gibertini2012}. There is experimental evidence,
based on the correlation between the carrier mobility and the width of the
resistance peak around charge neutrality, that the long-range disorder
potential (LRDP) due to SF could be the dominant source of disorder in
high-quality graphene on a substrate~\cite{Nuno14}. In addition, the SF
modulate the electron hopping energies between different atomic sites,
inducing the long-range disorder hopping (LRDH), leading to the appearance
of the (pseudo) vector potential~\cite{KatsnelsonBook,Vozmediano2010}.
Another common source of disorder are resonant scatterers (RS) such as
chemical species like hydrogen or organic groups, which also lead to a
sublinear carrier-density-dependent conductivity and a minimum conductivity
plateau around the neutrality point~\cite{NG10,WK10}.

Besides the transport properties, an important part of our knowledge about
the electronic properties derives from the optical spectroscopy measurements
\cite{Peres2010RMP,OP10}. Infrared spectroscopy experiments allow for the
control of interband excitations by means of electrical gating~\cite%
{WangF2008,LB08}. For doped pristine graphene with nonzero chemical
potential $\mu_F $, the optical conductivity is a step function $\sigma
\left( \omega \right) =\sigma _{0}\Theta \left( \omega -2\mu_F \right) $ at
zero temperature due to Pauli's exclusion principle. However, there are
experimentally observed background contributions to the optical spectroscopy
between $0<\omega <2\mu_F $~\cite{LB08,ChenCF2011}, which are due to the
extra intraband excitations introduced by disorder or many-body effects~\cite%
{Ando2002,Gruneis2003,Peres2006,Gusynin2006,SPG07,Gusynin2007,Stauber2008,Stauber2008b,MinHK2009,LB08,Mak2008,Yuan2011}%
. This opens the possibility to identify the source of disorder via the
optical measurements.

Previous theoretical investigation of disorders are mainly based on models
without considering the next-nearest-neighbor (NNN) hopping $t^{\prime }$.
The breakdown of electron-hole symmetry resulting from $t^{\prime }\not=0$
shifts the position of Dirac point from zero to $3t^{\prime }$~\cite%
{CG09,KatsnelsonBook}. Recent quantum capacitance measurements indicate that
the value of $t^{\prime }$ is about $0.3eV$~\cite{capacitance2013},
consistent with the values obtained from the density functional
calculations. It is generally thought that $t^{\prime }$ has relatively weak
effects on the physical properties of graphene at low energies~\cite%
{CG09,KatsnelsonBook,Stauber2008,capacitance2013}. %In this letter we show
%that including $t^{\prime }$ has a negligible effect in combination with
%long-range disorder such as CI, LRDP and LRDH, but changes the physics
%dramatically when RS are present. Different sources of disorder can be
%identified via their fingerprints in the transport and optical properties,
%and we use these fingerprints to demonstrate the dominant disorder source in
%several well-known experimental measurements.
%{\color{blue} text   }
In the present paper, we study the electronic, transport and optical
properties of graphene with different types of disorders including NNN. We
show that $t^{\prime }$ has a negligible effect in combination with
long-range disorder such as CI, LRDP and LRDH, but changes the physics
dramatically when RS are present. Different sources of disorder can be
identified via their fingerprints in the common measurable quantities, such
as dc conductivity, carrier mobility, optical spectroscopy and Landau level
spectrum \textit{\ etc.} We will use these fingerprints to demonstrate the
dominant disorder source in several well-known experimental measurements.
The paper is organized as follows. In Section II we gives a description of
the tight-binding Hamiltonian of single layer graphene including different
types of disorders and NNN. In section III{\ and }IV, we discuss the effect
of different disorders on the transport and optical properties of graphene.
Then, we study the Landau level spectrum and quantum capacitance in the
presence of perpendicular magnetic field in section V. Finally, a brief
discussion and conclusion, including a list of dominant disorder sources in
several experiments, is given in section VI.

\section{Model and Method}

%\label{Sec:Method}
%\textit{Model and Method---}
We consider disordered graphene described by the tight-binding (TB)
Hamiltonian
\begin{equation}
H=-\sum_{i,j}t_{i,j}c_{i}^{\dagger }c_{j}-\sum_{i,j}{t_{i,j}^{\prime }}%
c_{i}^{\dagger }c_{j}+\sum_{i}v_{i}c_{i}^{\dagger }c_{i},
\label{Hamiltonian0}
\end{equation}%
where the first sum is taken over nearest neighbors and the second one
is over next-nearest neighbors. %$v_{i}$ is the on-site potential.

For CI, we consider randomly distributed point-like charges at the center of
a hexagon of the honeycomb lattice ($\mathbf{r}_{k}$)\cite{Pereira2007},
which introduce the Coulomb energy $v_{i}=\sum_{k}\mathop{\mathrm{sign}}%
\left( k\right) e^{2}/\left( \kappa |\mathbf{r}_{i}-\mathbf{r}_{k}|\right) $
at each site $i$, and the screening effect due to the substrate is taken
into account by using the dielectric constant $\kappa $ of the substrate.
Here, according the values of $\mathrm{sign}\left( k\right) $ we consider
three types of CI: (1) CI$^{0}$ for randomly distributed positive or
negative potential caused by charges that the whole sample holds the
electric neutrality, (2) CI$^{+}$ for only positive potential and (3) CI$%
^{-} $ for only negative ones.

For LRDP, the on-site potential $v_{i}$ follows a corrected Gaussian profile
which varies smoothly on the scale of lattice constant as $%
v_{i}=\sum_{k}U_{k}\exp [-\left\vert \mathbf{r}_{i}-\mathbf{r}%
_{k}\right\vert ^{2}/(2d^{2})]$\cite{Yuan2011}, where $\mathbf{r}_{k}$ is
the $k-th$ Gaussian centers which are randomly distributed on the lattice
with probability $p_{v}$, $U_{k}$ represents the strength of the local
potential and is uniformly random in the range $[-\Delta _{v},\Delta _{v}],$
and $d$ is interpreted as the effective radius. We use $\Delta _{v}=t$ and $%
d=5a$ to represent the long-range Gaussian potential. Here $a\approx 1.42$%
\AA ~is the carbon-carbon distance in the single-layer graphene.

The LRDH is introduced in a similar way as LRDP except that the
nearest-neighbor hopping parameters are modified according a correlated
Gaussian form via $t_{ij}=t+\sum_{k}T_{k}\exp [-\left\vert \mathbf{r}_{i}+%
\mathbf{r}_{j}-2\mathbf{r}_{k}\right\vert ^{2}/\left( 8d_{t}^{2}\right) ],$%
where $T_{k}$, $d_{t}$ and $p_{t}\ $have similar meanings as in LRDP, and we
choose $\Delta _{t}=0.25t$ and $d_{t}=5a$\cite{Yuan2011}. We want to
emphasis that, although the amplitude ($\Delta $) and radius ($d$) of the
Gaussian profile in the LRDH and LRDP are free parameters that can be turned
in the tight-binding model, the numerical results show little quantitative
difference as long as these parameters are of the same order as the chosen values. In
general, an increase (decrease) of the amplitude or radius is equivalent to an
increase (decrease) of the disorder concentration.

The hydrogen-like RS is described by the Hamiltonian $H_{RS}=V\sum_{i}\left(
d_{i}^{\dagger }c_{i}+\mathrm{h.c}\right) $~\cite{Robinson08,WK10,YRK10},
where $V$ is the hopping between carbon and adatom. We consider the limiting
case with $V\rightarrow \infty $, i.e., the electron at the impurity site is
completely localized such that the resonant scatterer behaves like vacancy~%
\cite{WK10}. In our calculations, we use $t\approx 2.7$ eV and ${t^{\prime }}%
=t/10$ for the nearest and next-nearest neighbor hopping parameters,
respectively. The spin degree of freedom contributes only through a
degeneracy factor and, for simplicity, has been omitted in Eq.~(\ref%
{Hamiltonian0}).

The calculations of the electronic and optical properties are performed by
the tight-binding propagation method (TBPM)~\cite{HR00,YRK10,WK10,Yuan2012},
which is based on the numerical solution of the time-dependent Schr\"{o}%
dinger equation and Kubo's formula. The advantage of this method is that all
the calculated quantities are extracted from the real-space wave propagation
without any knowledge of the energy eigenstates. Furthermore one can
introduce different kinds of (random) disorder by constructing the
corresponding TB model for a sample scaling up to micrometers. For more
details about the numerical methods we refer to Refs~%
\onlinecite{YRK10,Yuan2011}. The simulated graphene sample contains up to $%
8192\times 8192$ atoms subject to periodic boundary conditions.

%\emph{Transport properties }---

\section{TRANSPORT PROPERTIES}

We first consider the carrier-density-dependence of the microscopic
conductivity $\sigma \left( n_{e}\right) $ for disordered graphene. The
microscopic (or semi-classic) conductivity is calculated from the diffusive
region of the charge transport, i.e., when the time-dependent diffusion
coefficient reaches its the maximum \cite{Cresti2013,Laissa2013,Laissa2014},
and it is comparable to the conductivity extracted from the field-effect
measurements. In TBPM, the microscopic conductivity at an energy $E$ is
calculated by using the Kubo formula \cite{YRK10,Yuan2012}
\begin{eqnarray}
\mathbf{\sigma }\left( E\right)  &=&\underset{\tau }{\max }\frac{\rho \left(
E\right) }{\Omega }\int_{0}^{\tau }dt~\text{Re}\left[ e^{-iEt}\left\langle
\varphi \right\vert Je^{i\mathcal{H}t}J\left\vert E\right\rangle \right] ,
\notag \\
&&
\end{eqnarray}%
where $\left\vert \varphi \right\rangle $ is a normalized random state, $%
\left\vert E\right\rangle $ is the \textit{normalized} quasi-eigenstate\cite%
{YRK10}, $J$ is the current operator, $\Omega $ is the sample area, and $%
\rho \left( E\right) $ is the density of states (DOS) calculated via\cite%
{HR00,YRK10}
\begin{equation}
\rho \left( E\right) =\frac{1}{2\pi }\int_{-\infty }^{\infty
}e^{iEt}\left\langle \varphi |\varphi (t)\right\rangle dt.  \label{Eq:DOS}
\end{equation}%
The measured field-effect carrier mobility is related to the microscopic
conductivity as $\mu \left( E\right) =\sigma \left( E\right) /en_{e}\left(
E\right) $, where the carrier density $n_{e}$ is obtained from the integral
of DOS via $n_{e}\left( E\right) =\int_{0}^{E}\rho \left( \varepsilon
\right) d\varepsilon $.

From the results shown in Fig.~\ref{Fig:DC}, we see that (1) including $t^{\prime }$
has  negligible effects for CI, LRDP and LRDH, but the results for RS change
dramatically. In the presence of RS, there is a strong electron-hole
asymmetry in the carrier-density-dependence of dc conductivity. This is due
to the fact that the impurity band created by RS is shifted from the Dirac
point to the hole side\cite{Pereira2008}, introducing strong electron-hole
asymmetry at low energies; (2) as a consequence of this shift the
conductivity plateau around the neutrality point is also shifted to the hole
side, with an impurity-concentration dependent height and width (for very
small concentration of RS, there is just as a kink instead of a plateau, see
the point indicated by an arrow in Fig.~\ref{Fig:DC}(h)) ; %{\color{blue}
These features can be observed in graphene if the
concentration of generic RS is increased by exposing the material to atomic hydrogen \cite%
{NG10}. (3) $\sigma \left( n_{e}\right) $ exhibits a sublinear dependence
for small concentration for all types of disorders, except for the hole side
in the presence of RS; (4) For LRDH, $\sigma \left( n_{e}\right) $ is
insensitive to the changes of the disorder concentration ($p_{t}$); (5) No
matter whether $t^{\prime }$ is nonzero or not, linear-dependent $\sigma
\left( n_{e}\right) $ appears only in CI with large concentration of $n_{C}$
\cite{ZhuW2013}, indicating that CI is the dominant source of disorder in
the
% concentration $n_{C}> 0.01\%$, indicating that CI is the dominant disorder source for
experimental samples which show clearly the linear carrier-density-dependent
conductivity (such as K151 in Ref. \cite{Tan2007}, and Potassium doped
samples in Ref. \cite{ChenJH2008}, etc.), agree with the theoretical
prediction that $\sigma \left( n_{e}\right) \propto $ $n_{e}$; (6) The
electron-hole asymmetry appears also for larger concentration of CI if there
is only one types of charge resource (CI$^{+}$ and CI$^{-}$). However, this
asymmetry is different from the one due to RS in two aspects: first,
for CI there is no kink or plateau in the profile; second, the
conductivities on both electron and hole sides decrease significantly with
larger concentration of CI; (7) Only in the case of CI$^{+}$ the conductivity
on the electron side is smaller than on the hole side with the same
concentration of carrier density, which is a unique signature of CI$^{+} $.
This is in concert with experiment results \cite{Tan2007,ChenCF2011}.
% consistent with previous transport calculations on CI\cite{Ando2006,Hwang2007,Pereira2007,ZhuW2013}.

The field-effect carrier mobility $\mu $ can be calculated from the
conductivity and carrier-density through $\mu =\sigma /en_{e}$. In the following
we show only the results with non-zero $t^{\prime }$. From the results presented in
Fig.~\ref{Fig:mobility}, we see that (1) the carrier-dependence of mobility $%
\mu \left( n_{e}\right) $ is very similar for CI$^{0}$ and LRDP; (2) for
LRDH, $\mu \left( n_{e}\right) $ is insensitive to the disorder strength;
(3) electron-hole asymmetry appears for CI$^{+}$, CI$^{-}$ and RS, but only
in the case of CI$^{+}$ the electron mobility is smaller than the hole for
the same concentration of carrier density; (4) for RS, the mobility on the
electron side is insensitive to the impurity concentration, and its value
can be one order of magnitude larger than the value on the hole side. For
example, considering a RS concentration of $n_{x}=0.025\%$, the electron
mobility at carrier density $5\times 10^{12}cm^{-2}$ is about $3,000$ ($%
cm^{2}V^{-1}s^{-1}$) but the hole mobility for the same carrier density is
only $\sim 300$. This significant one order difference of the electron and
hole mobility is a unique signature of RS; (5) with RS present, on the hole
side, the carrier-density-dependent mobility is not monotonic and $\mu
\left( n_{e}\right) $ reaches a minimum at the density corresponding to the
tail of the conductivity plateau. However with RS present and $t^{\prime }=0$%
, the drop of mobility at the minimum is one order of magnitude larger than
the experimental result.

%====================================
\begin{figure}[t]
\begin{center}
\mbox{
 \includegraphics[width=0.48\columnwidth]{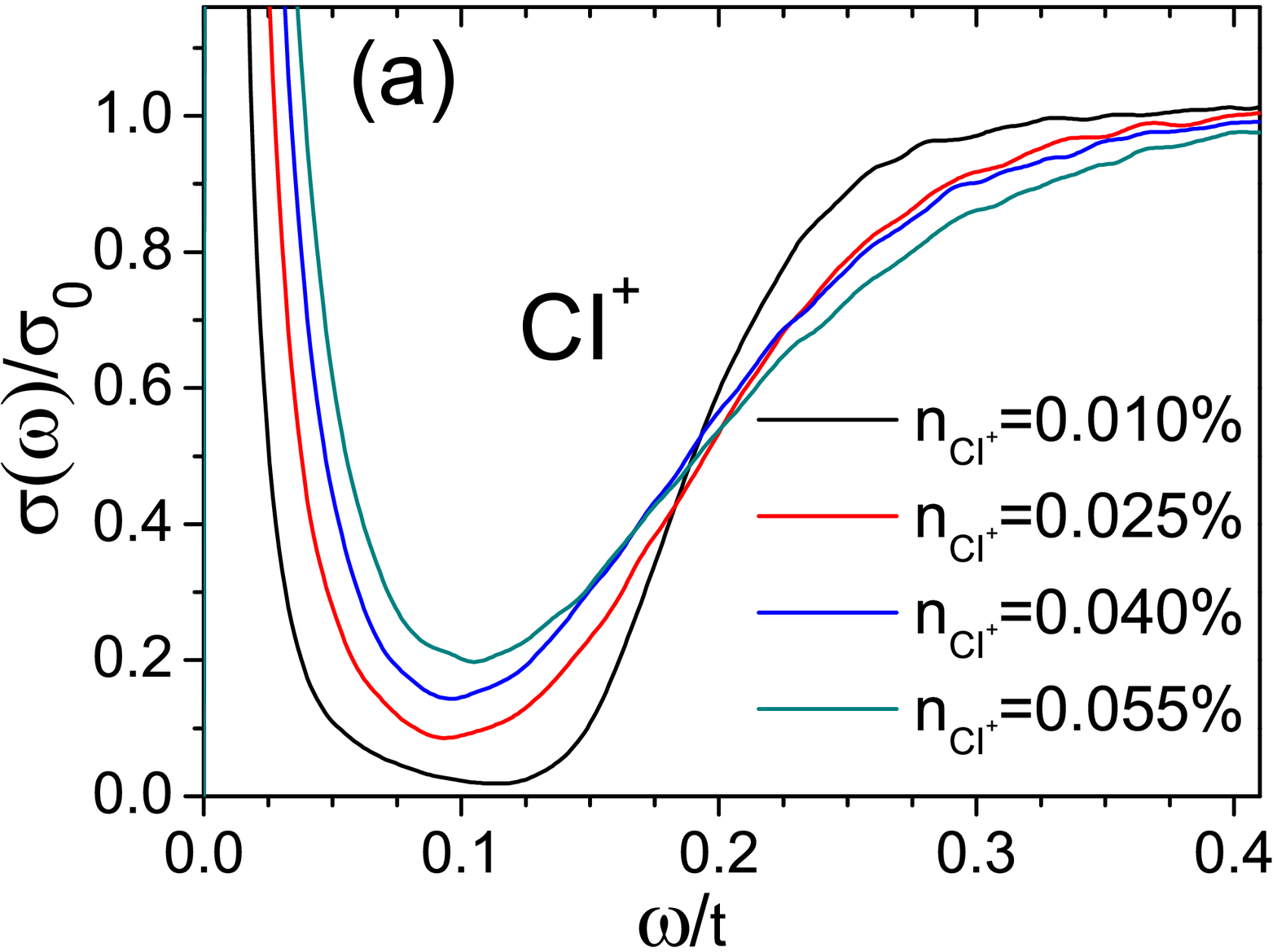}
 \includegraphics[width=0.48\columnwidth]{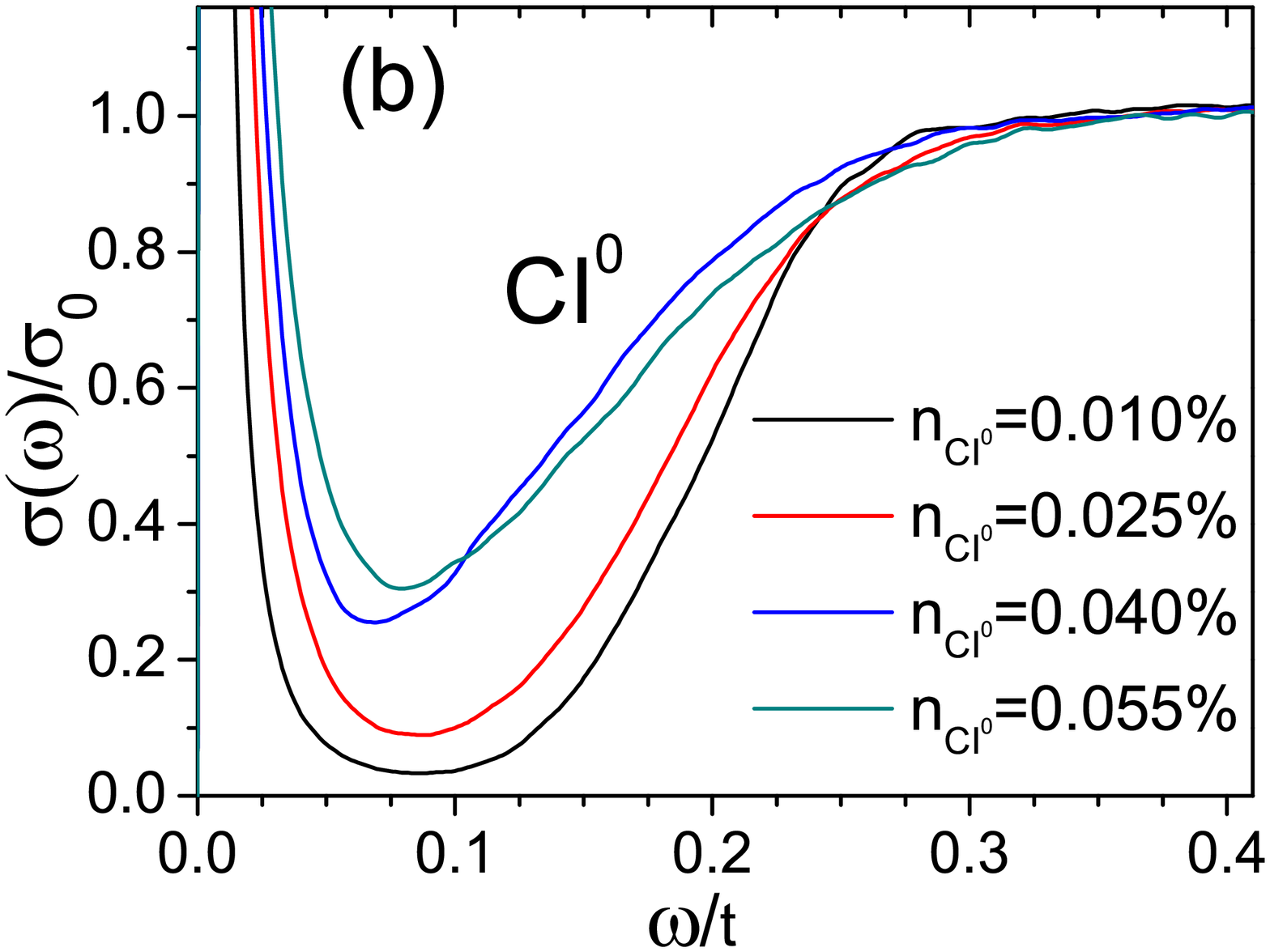}
  }
\mbox{
 \includegraphics[width=0.48\columnwidth]{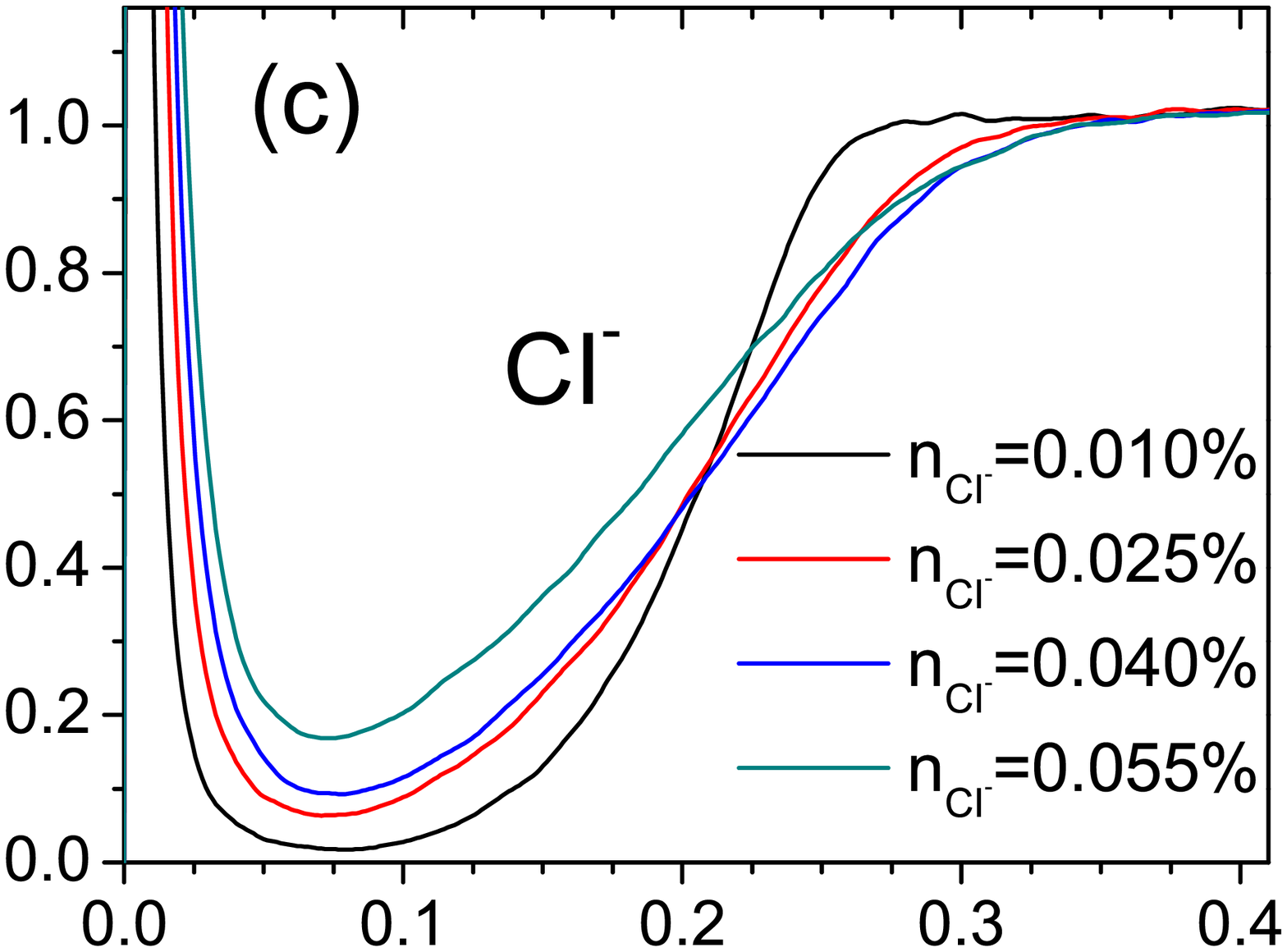}
 \includegraphics[width=0.48\columnwidth]{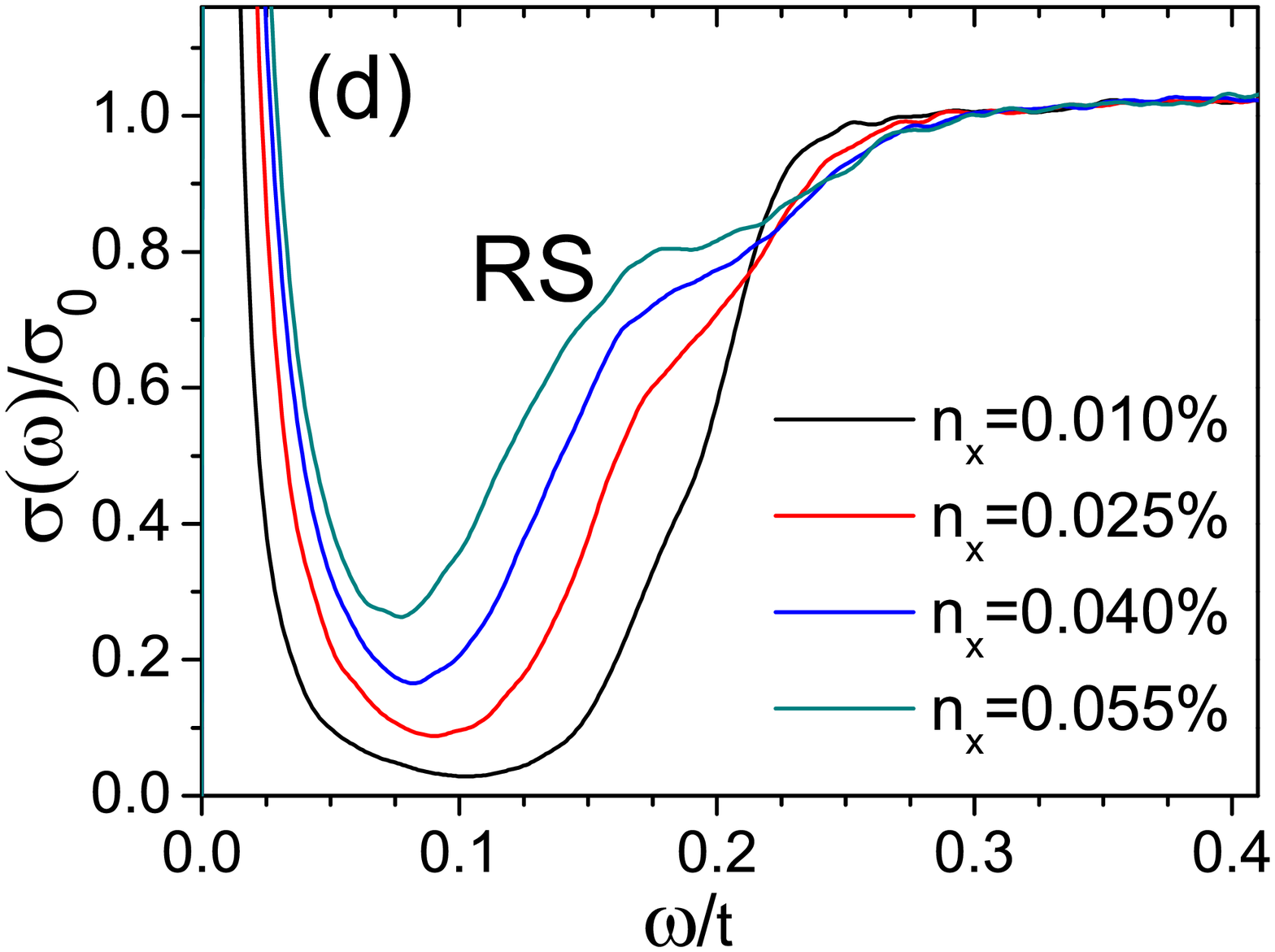}
  }
\mbox{
 \includegraphics[width=0.48\columnwidth]{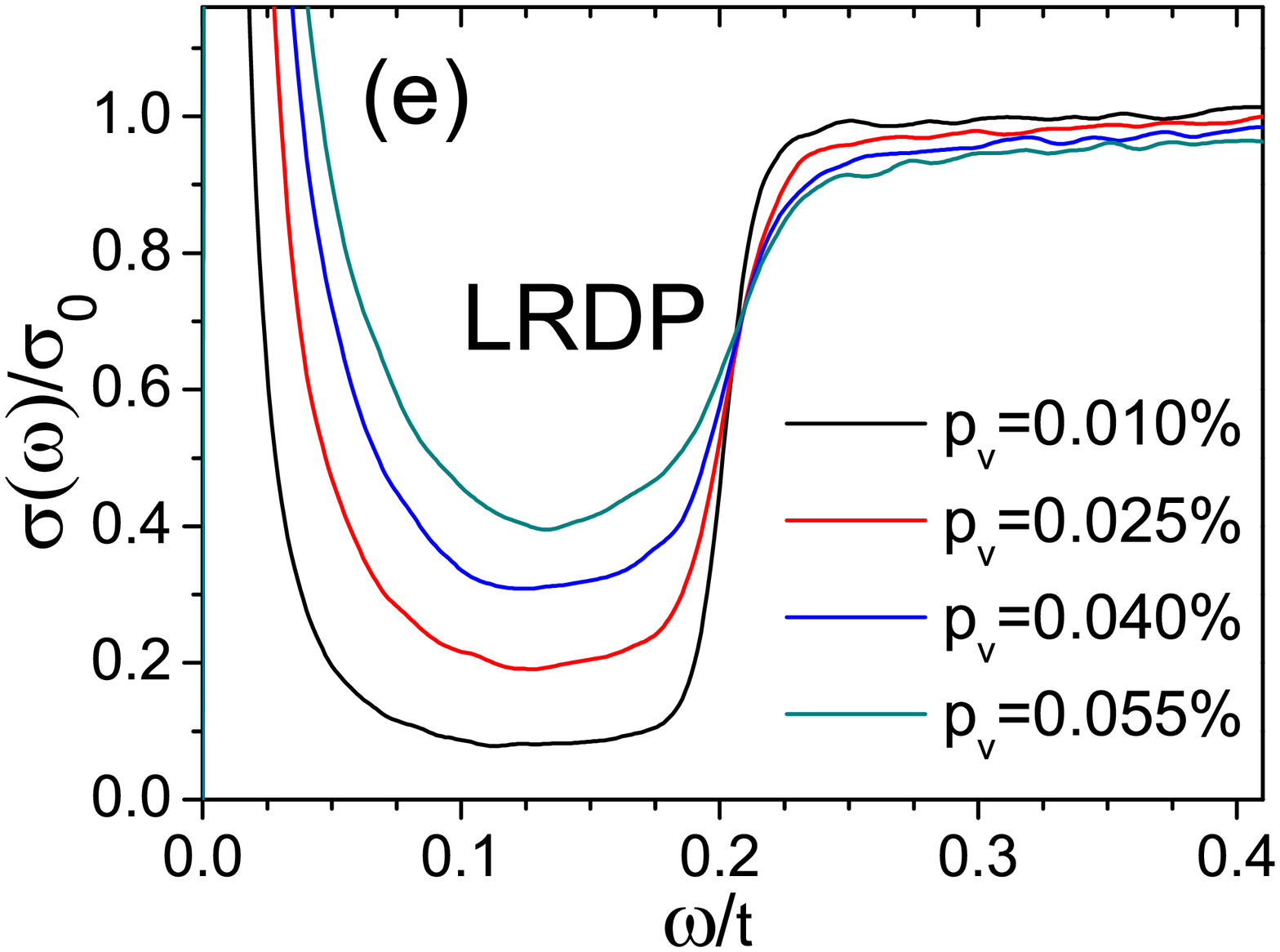}
 \includegraphics[width=0.48\columnwidth]{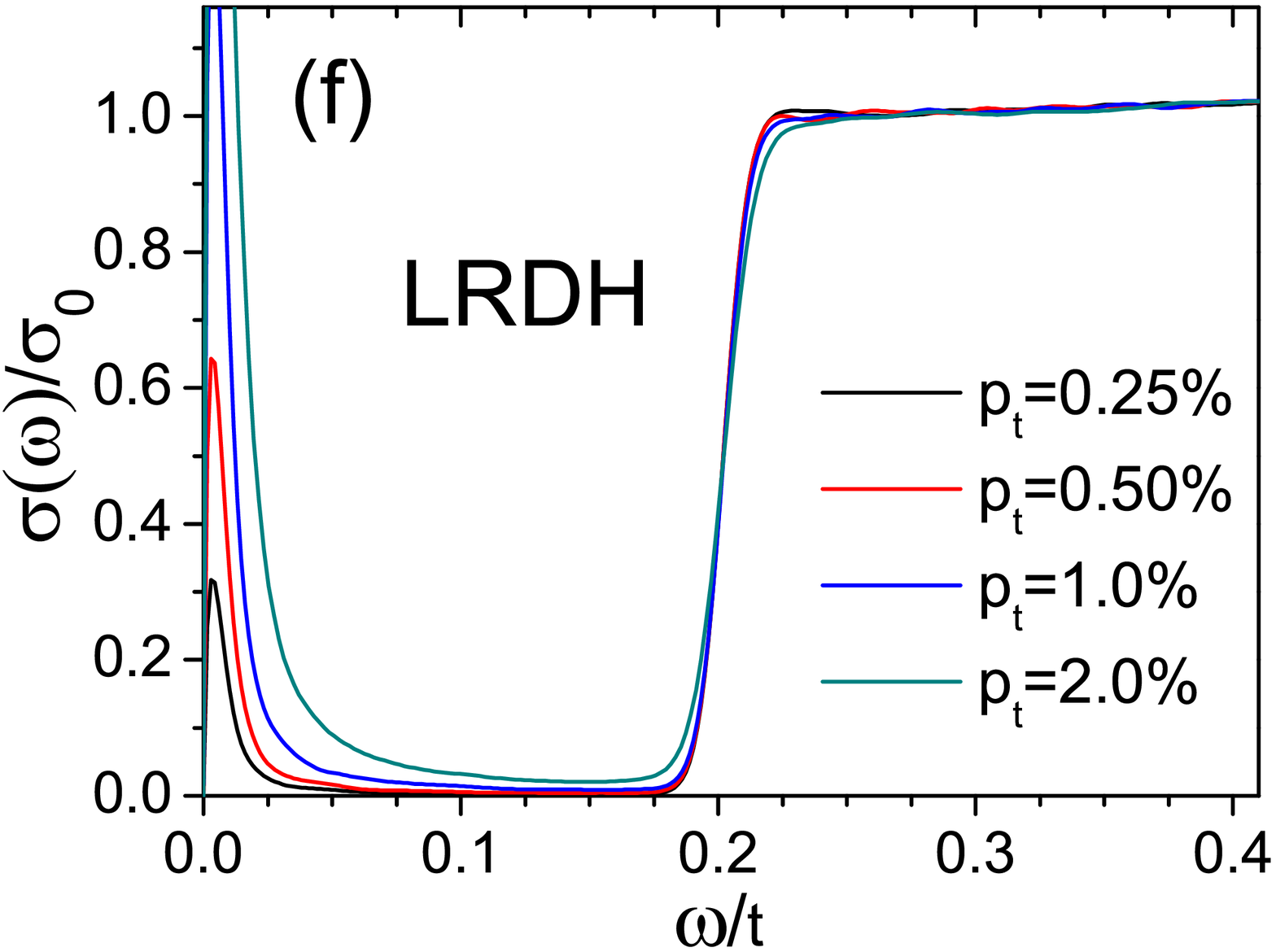}
  }
\end{center}
\caption{ (Color online) The optical conductivity as a function of energy
for disordered graphene with $\protect\mu_F=0.1t$ and $t^{\prime }=0.1t$.
Here $\sigma _{0}=\pi e^{2}/(2h)$ is the universal optical conductivity of graphene.
All along the work the temperature of optical calculation is $T=45K$, the
same as in the experiment of Ref.\onlinecite{LB08}.}
\label{Fig:ACTprime}
\end{figure}

\begin{figure*}[tb]
\includegraphics[width=0.8\linewidth]{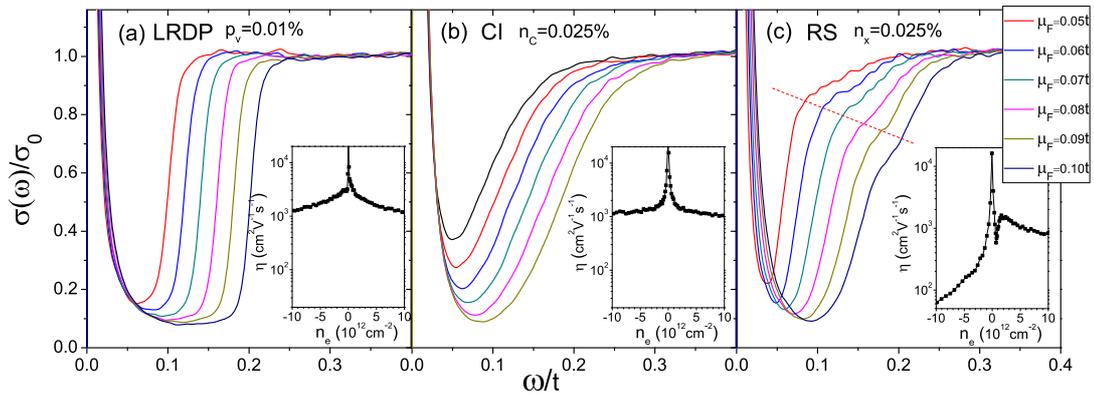} %[tph]
\caption{ (Color online) The optical conductivity as a function of energy
for disordered graphene (a) LRDP, (b) CI, and (c) RS with $t^{\prime }$. The
disorder concentrations are determined via the best fit to the experimental
results of optical spectroscopy~\protect\cite{LB08,ChenCF2011}. The chemical
potential $\protect\mu_F $ changes from $0.05t$ to $0.1t$. The inner panels
show the corresponding carrier mobility for the same concentration of
disorder. The dashed line in (c) is a guide to eye separating two region in
which the spectrum changes differently in the presence of RS.}
\label{Fig:AC}
\end{figure*}

The minimum conductivity $\sigma _{\min }$ at the Dirac point is of the
order of $4e^{2}/h$ for all types of long-range disorders with $t^{\prime
}=t/10$. The values of $\sigma _{\min }$ in CI and LRDP do not depend on $%
t^{\prime }$, but change with the disorder strength such that larger
concentration of disorder leads to larger values of $\sigma _{\min }$. This
is due to the fact that the increase of potential sources in CI and LRDP
will increase the DOS at the $\mu _{F}$, leading to more states which can
contribute to the transport. This may also explain the experimental
observations in Ref.~\onlinecite{Tan2007} and Ref.~\onlinecite{Geim2007} in
which the low mobility does not necessary correspond to a smaller value of $%
\sigma _{\min }$. For LRDH, the value of $\sigma _{\min }$ for $t^{\prime
}=0 $ is about two time larger than the value for $t^{\prime }=t/10$, but
both are insensitive to the disorder strength. For RS and $t^{\prime }=0$, $%
\sigma _{\min }$ is of the order of $4e^{2}/\pi h$, independent on the
impurity concentration $n_{x}$~\cite{Cresti2013,Laissa2013,Laissa2014}, but
if $t^{\prime }=t/10$, $\sigma _{\min }$ from being of the order of $4e^{2}/h$
at small $n_{x}$ to $4e^{2}/\pi h$ when $n_{x}\geq 0.1\%$, consistent with
the numerical results of Ref.~\onlinecite{Laissa2014} (data not shown). Thus
we conclude that our results indicate that the minimum conductivity $4e^{2}/h
$ found in the experiments is dominated by long-range disorder but that the
value of $4e^{2}/\pi h$ is due to RS only. It is worth to mention that our consideration does not take into
account the effects of weak (anti)localization which can change the behavior
of conductance at very large distances~\cite{Andersontransitions2008}, due to energy smearing
in our calculations. The latter works as dephasing. At the same time,
this dephasing can be physical for real samples.

\section{OPTICAL SPECTROSCOPY}

%\emph{Optical properties }---

The optical conductivity is calculated by using the Kubo formula \cite%
{Ishihara1971} within TBPM\cite{YRK10} as (omitting the Drude contribution
at $\omega =0$)
\begin{eqnarray}
\sigma \left( \omega \right) &=&\lim_{\epsilon \rightarrow 0^{+}}\frac{%
e^{-\beta \omega }-1}{\omega \Omega }\int_{0}^{\infty }e^{-\epsilon t}\sin
\omega t  \notag  \label{gabw2} \\
&&\times 2~\text{Im}\left\langle \varphi |f\left( \mathcal{H}\right) J\left(
t\right) \left[ 1-f\left( \mathcal{H}\right) \right] J|\varphi \right\rangle
dt,  \notag \\
&&  \label{Eq:OptCond}
\end{eqnarray}%
where $\beta =1/k_{B}T$ is the inverse temperature, $f\left( \mathcal{H}%
\right) =1/\left[ e^{\beta \left( \mathcal{H}-\mu _{F}\right) }+1\right] $
is the Fermi-Dirac distribution operator. Similar as for the transport
properties, our numerical calculations show that $t^{\prime }$ has
negligible effects on the optical properties of disordered graphene, except
if RS are present. In general, disorder introduces new states which could
contribute to the extra intraband excitations~\cite%
{Ando2002,Gruneis2003,Peres2006,Gusynin2006,SPG07,Gusynin2007,Stauber2008,Stauber2008b,MinHK2009,LB08,Mak2008,Yuan2011}%
, and therefore enhances the optical conductivity below $2\mu _{F}$, which
might explain the observed background contribution in the optical spectrum
for $0<\omega <2\mu _{F}$~\cite{LB08,ChenCF2011}. This is confirmed by the
optical conductivity of disordered graphene calculations shown in Fig.~\ref%
{Fig:ACTprime}. For disordered graphene with CI (including CI$^{0}$, CI$^{+}$
and CI$^{-}$) there is a strong enhancement of the optical conductivity
below $2\mu _{F}$ and the enhanced spectrum forms a plateau with
disorder-dependent minimum conductivity. For LRDP, there is in addition a
disorder-dependent plateau in the optical spectrum below $2\mu _{F}$, which
is much wider that the one due to CI. For LRDH, the enhancement of the
optical conductivity is much smaller than for other types of disorders. For
RS and $t^{\prime }=0$, a disorder-dependent peak appears at $\omega \approx
\mu _{F}$, which is due to the enhanced excitations of the midgap states at
the Dirac point. This peak disappears for $t^{\prime }=t/10$, and instead, a
disorder-dependent narrow plateau appears.

In practice, instead of varying the disorder concentration, it is easier to
change the chemical potential by applying an electrical potential to a gate.
In order to compare to the experimental data of the spectroscopy
measurements~\cite{LB08,ChenCF2011} quantitatively, we plot in Fig.~\ref%
{Fig:AC} the best fit of the optical conductivity for different chemical
potentials ranging from $0.05t$ to $0.1t$ (since the results of CI$^{0}$, CI$%
^{+}$ and CI$^{-}$ are similar, we present here only the case of CI$^{0}$).
The disorder concentrations shown in Fig.~\ref{Fig:AC} are determined by
matching the minimum value of the optical conductivity plateau to the one
observed~\cite{LB08,ChenCF2011}, yielding $\sigma _{plateau}$ of the order
of $0.1\sigma _{0}$ for $\mu _{F}\approx t/10$. The best match of the
disorder concentrations from our simulations is $p_{v}=0.01\%$ for LRDP, $%
n_{C}=0.025\%$ for CI and $n_{x}=0.025\%$ for RS. A direct comparison of the
profile of the spectrum between our simulations and the experiments in Ref.~%
\onlinecite{LB08,ChenCF2011} indicates that LRDP fits best to the
experiments. In Ref.~\onlinecite{LB08}, the carrier mobility measured for
the same device is as high as $8,700cm^{2}V^{-1}s^{-1}$ at carrier densities
of $2\times 10^{12}cm^{-2}$, and the LRDP also gives the highest mobility
that it can reach $\sim 3,000$. For CI, $\mu \sim 1500$, and for RS the
mobility is even smaller: for electrons it is $\sim 1,000$ and for holes $%
\sim 300$. Therefore we conclude that the background contribution of the
optical conductivity below $2\mu _{F}$ as observed in Ref.~\onlinecite{LB08}
should be due mainly to the presence of LRDP.

%________________________________________________
\begin{figure*}[t]
\mbox{
 \includegraphics[width=6cm]{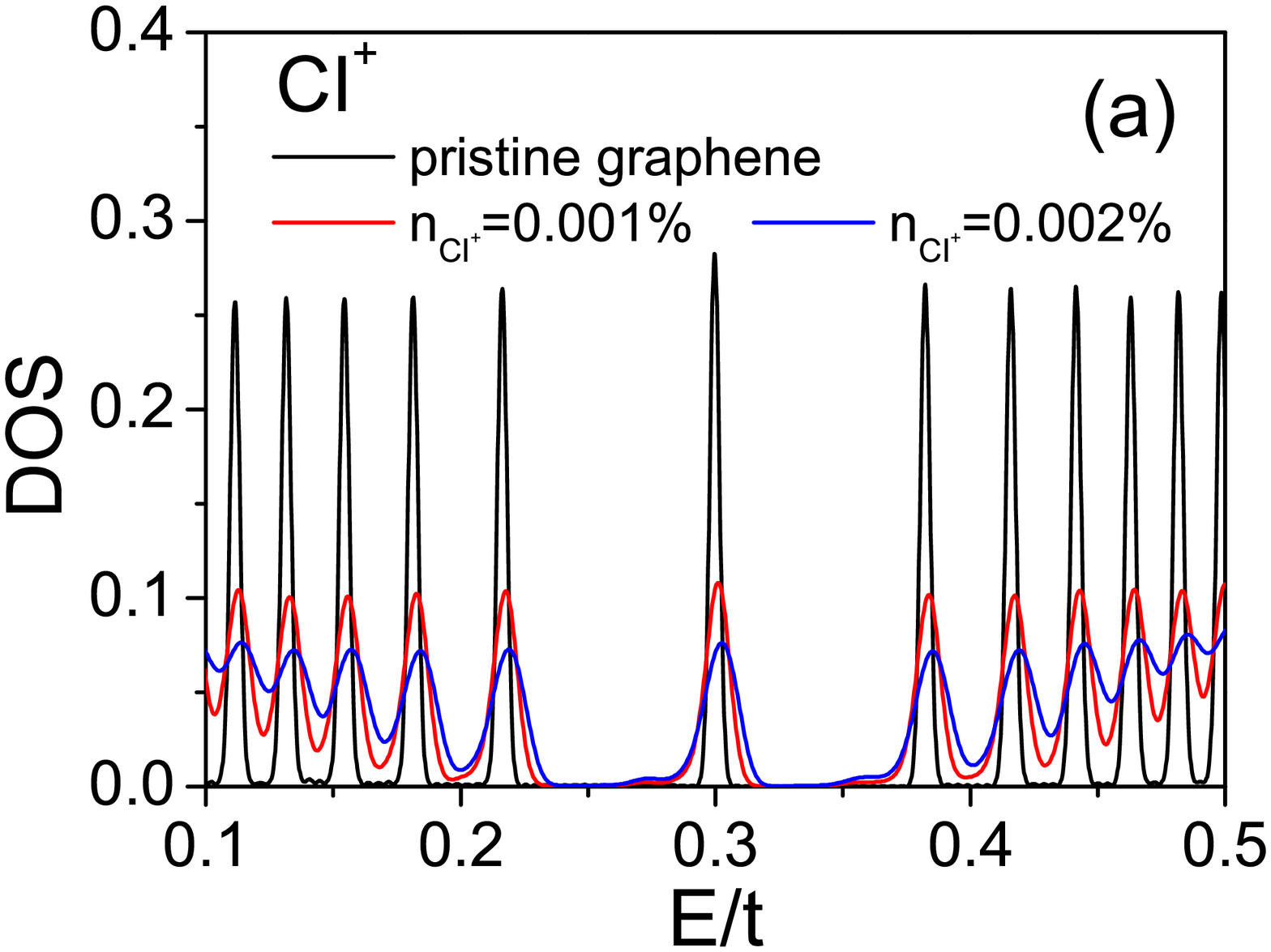}
 \includegraphics[width=6cm]{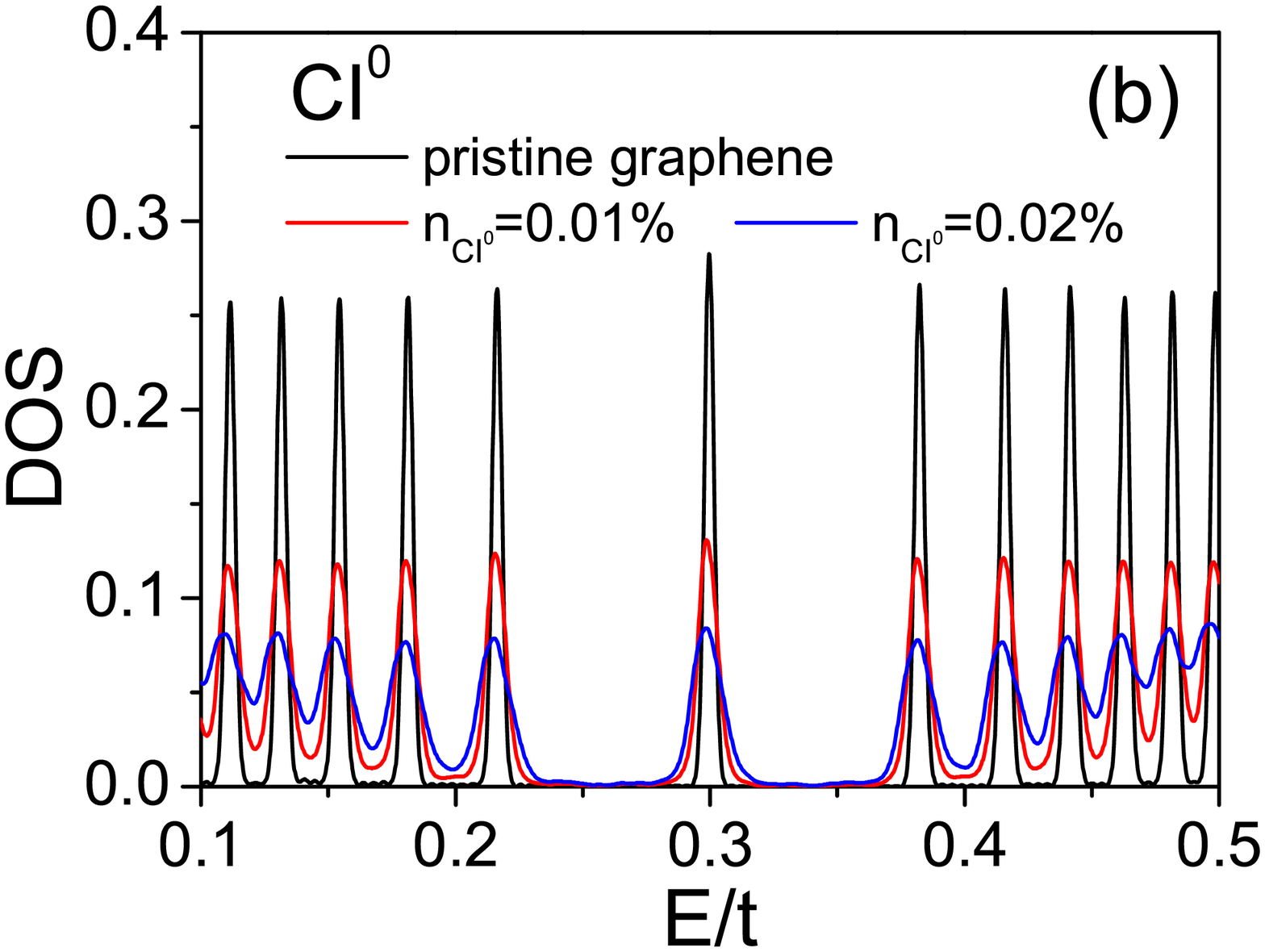}
  }
\mbox{
 \includegraphics[width=6cm]{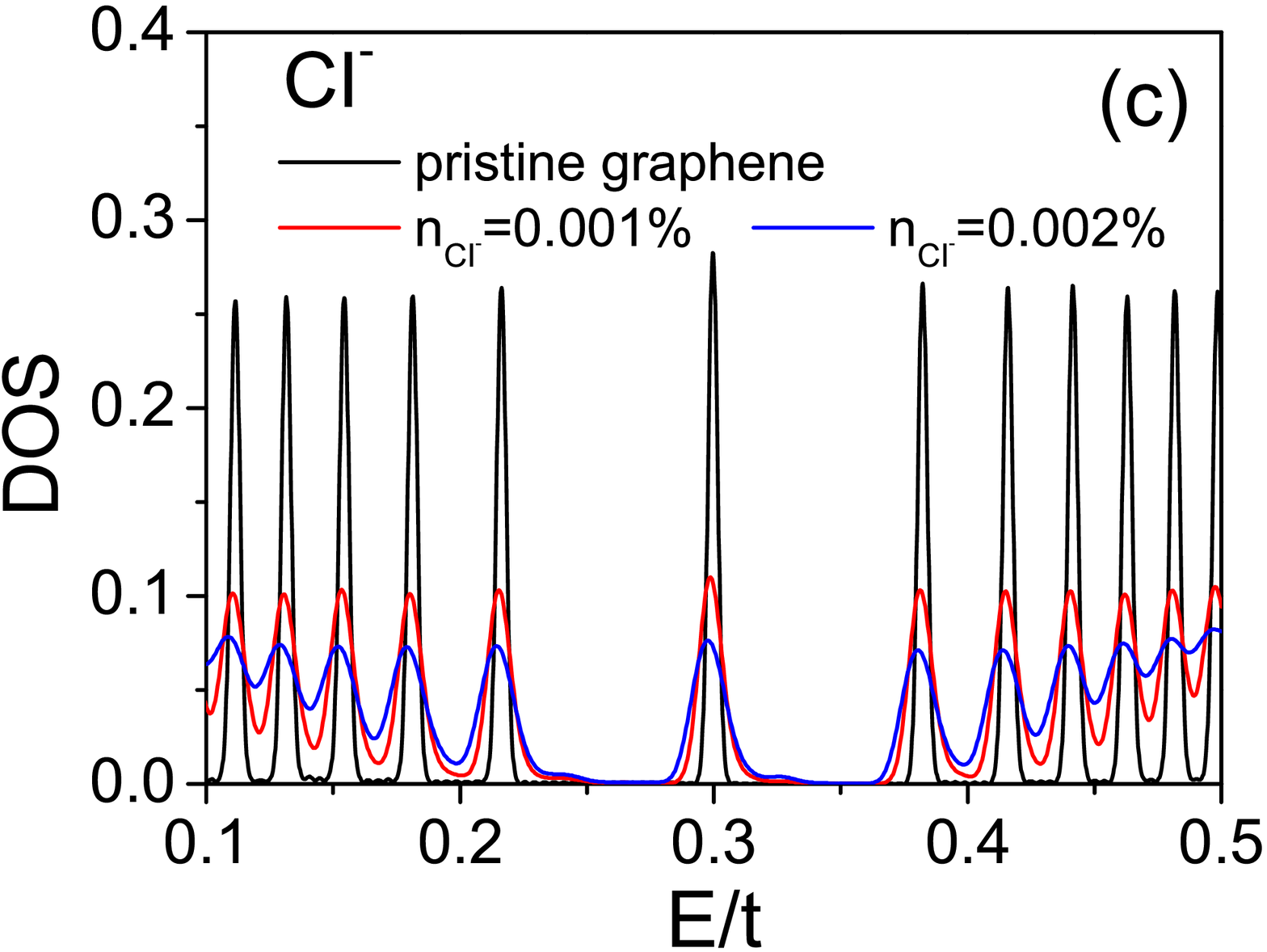}
 \includegraphics[width=6cm]{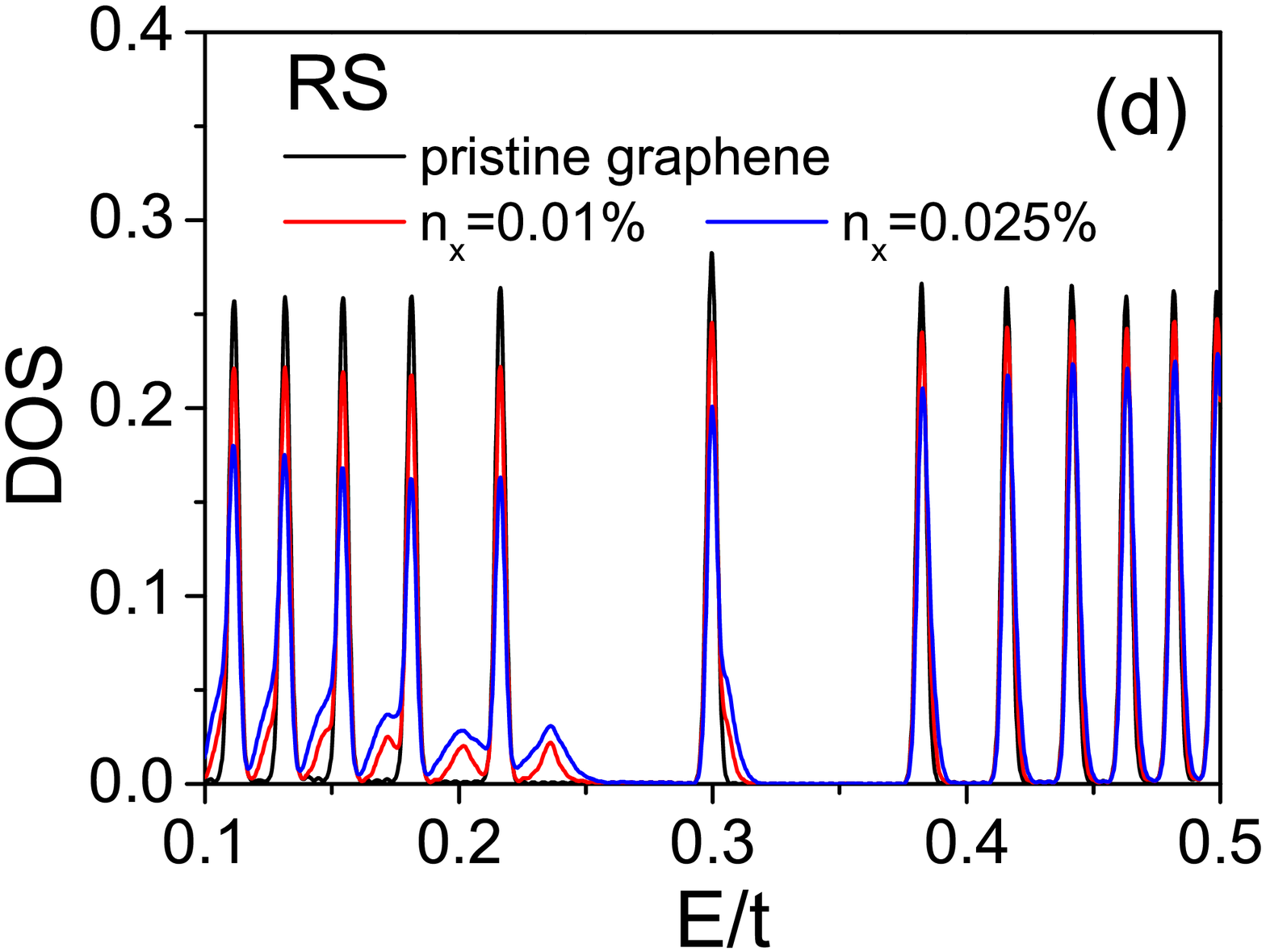}
  }
\mbox{
 \includegraphics[width=6cm]{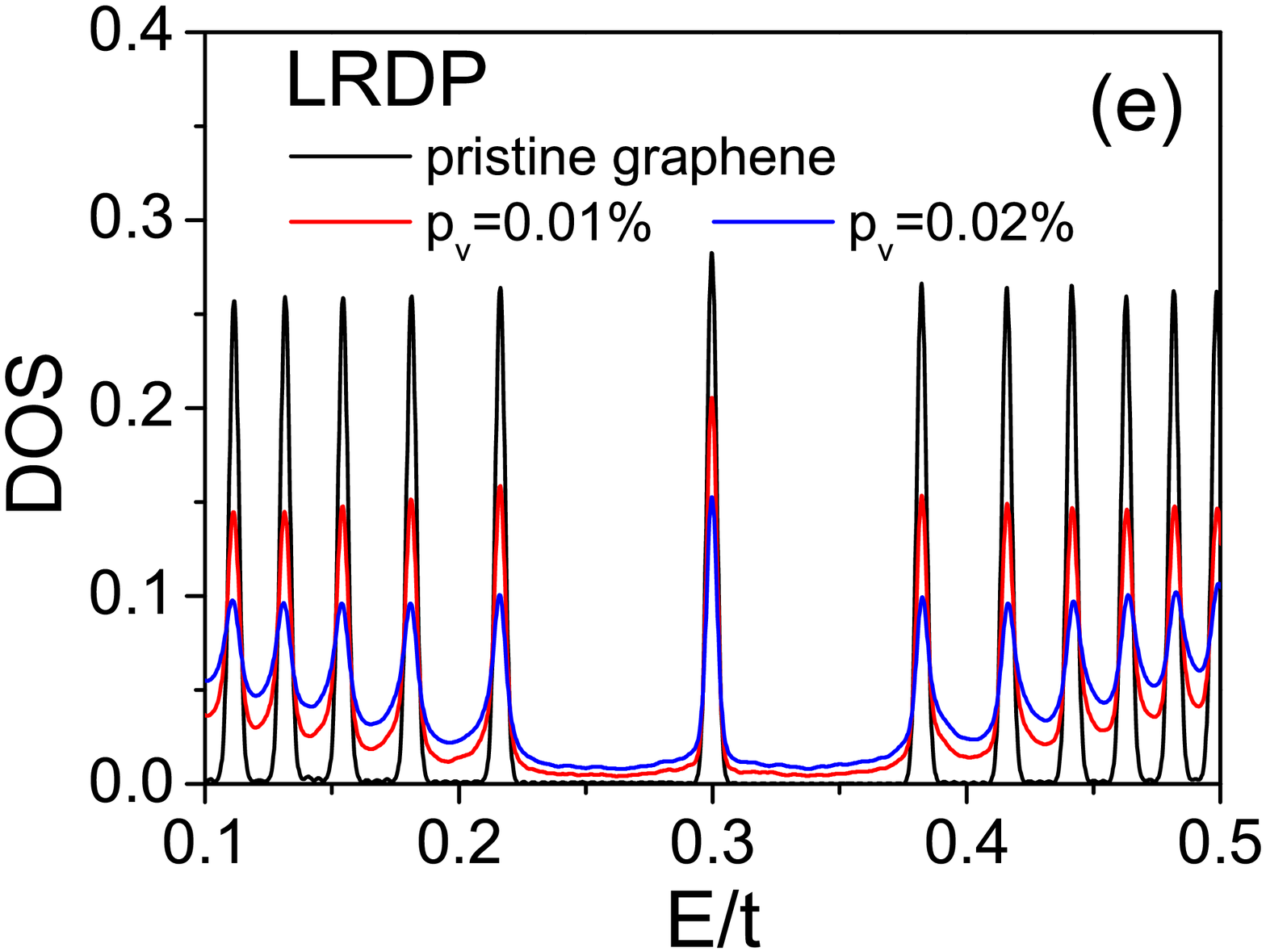}
 \includegraphics[width=6cm]{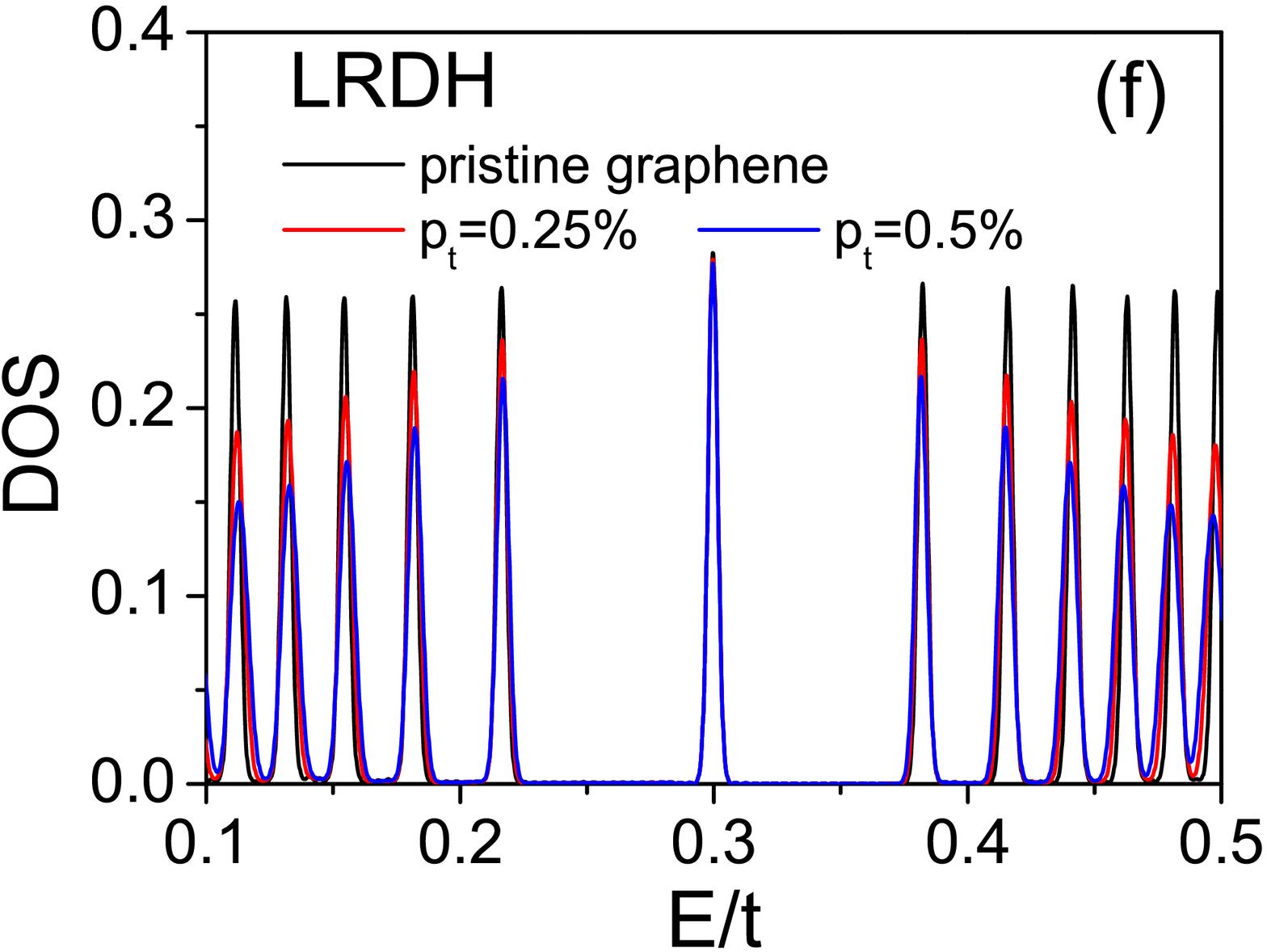}
  }
\caption{ (Color online) Density of states as a function of energy for
disordered graphene in the presence of a uniform perpendicular magnetic
field ($B=50$T).}
\label{Fig:DOS}
\end{figure*}
%________________________________________________

%________________________________________________
\begin{figure*}[t]
\mbox{
 \includegraphics[width=6cm]{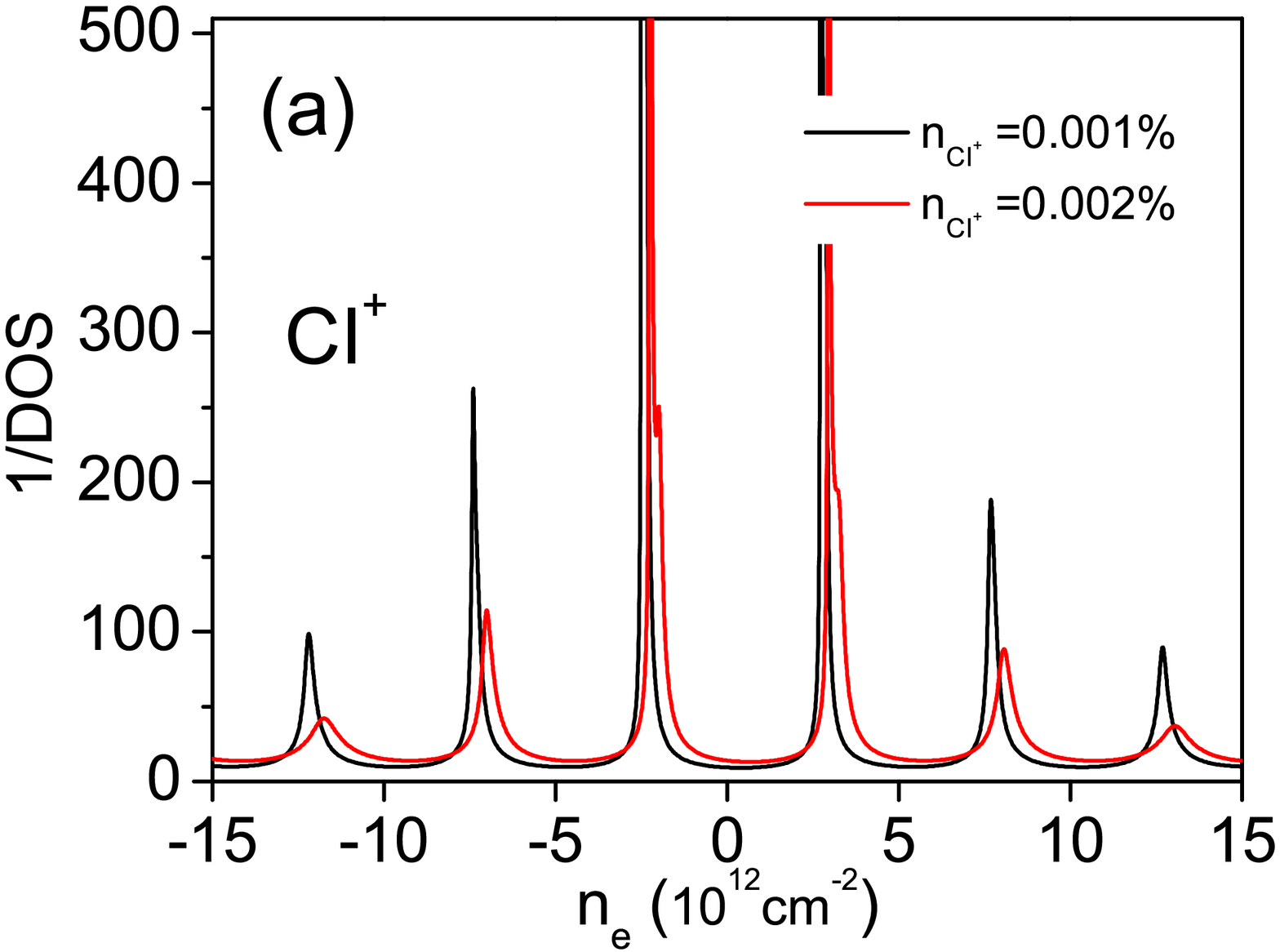}
 \includegraphics[width=6cm]{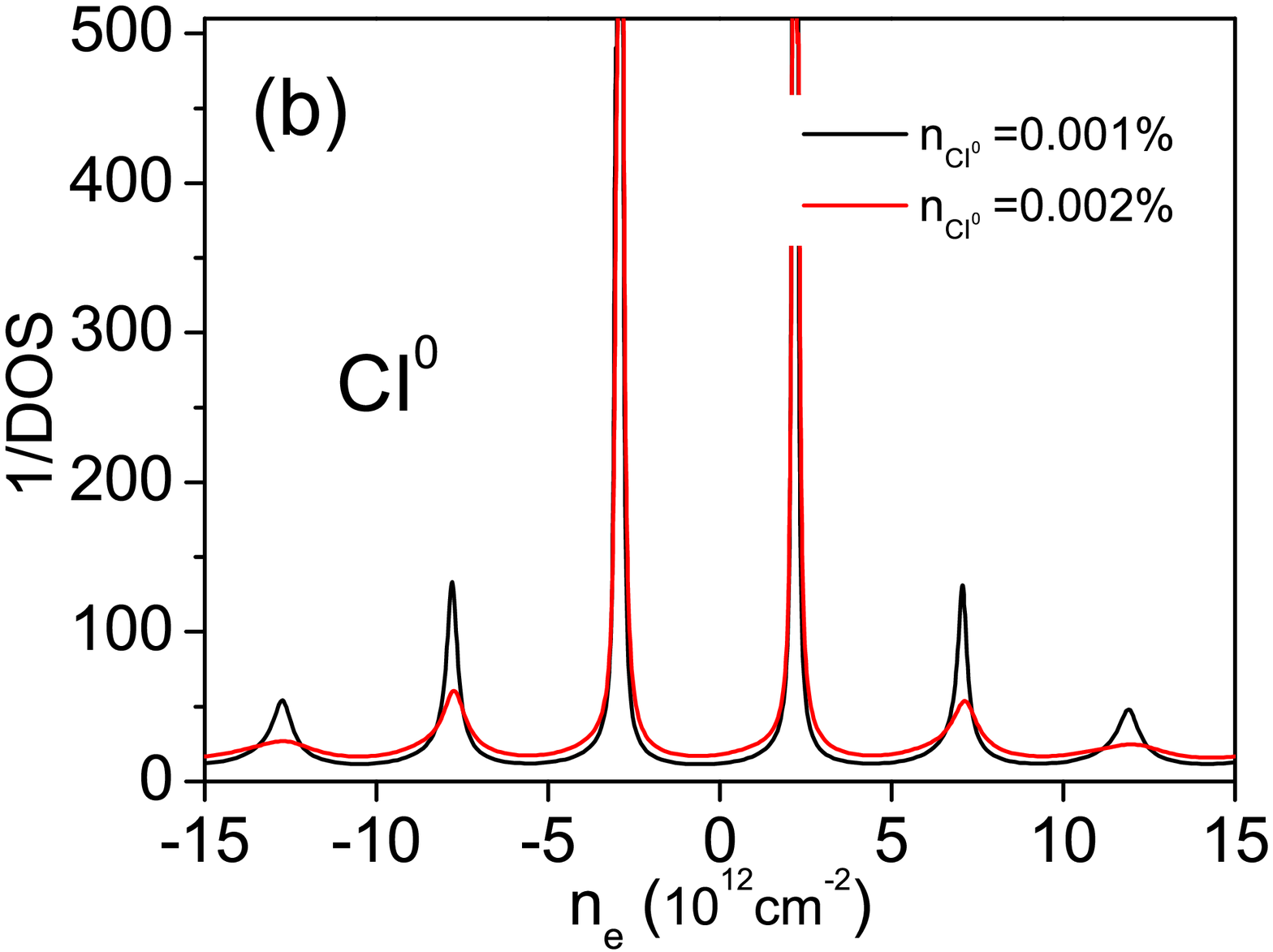}
  }
\mbox{
 \includegraphics[width=6cm]{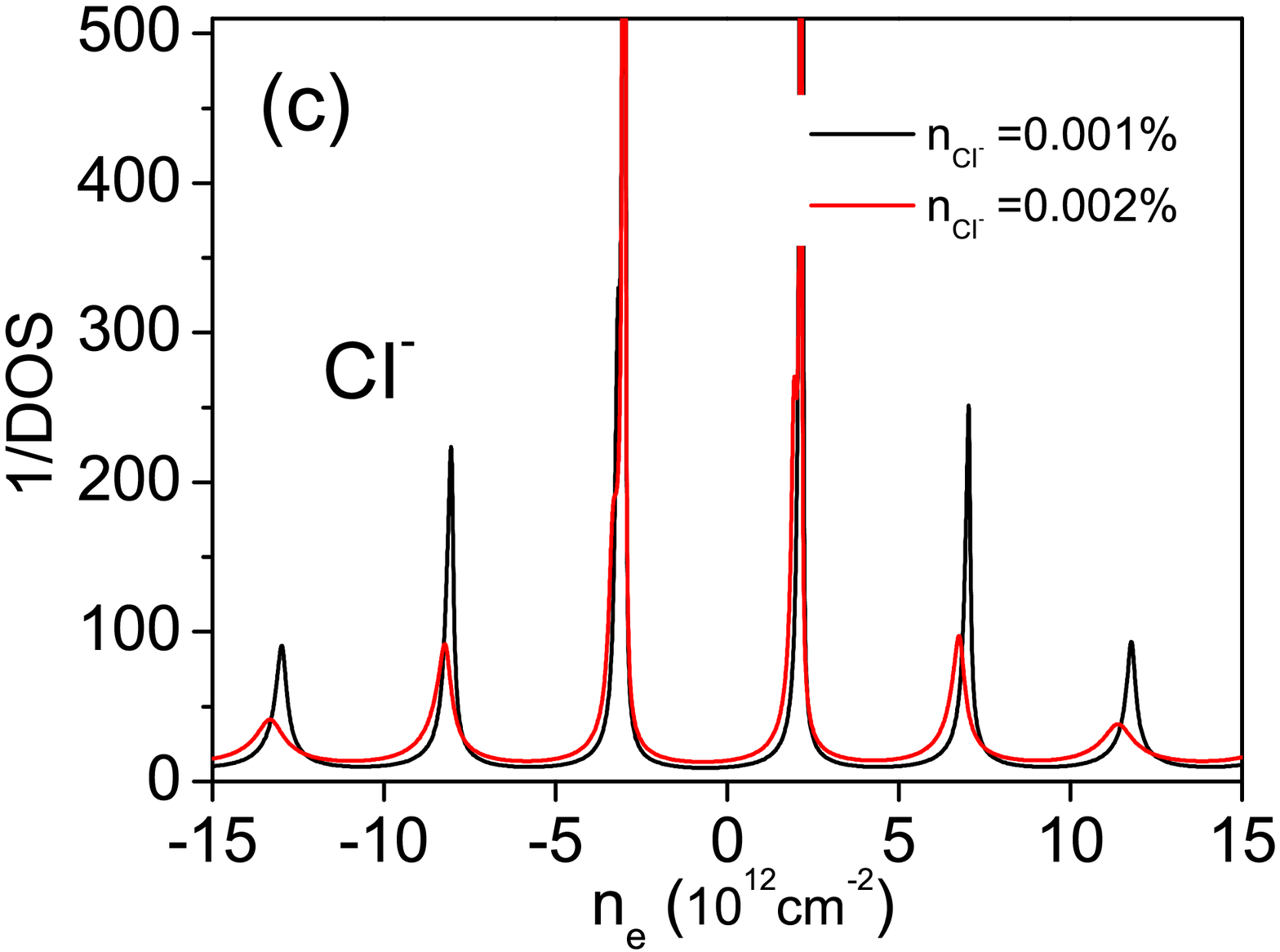}
 \includegraphics[width=6cm]{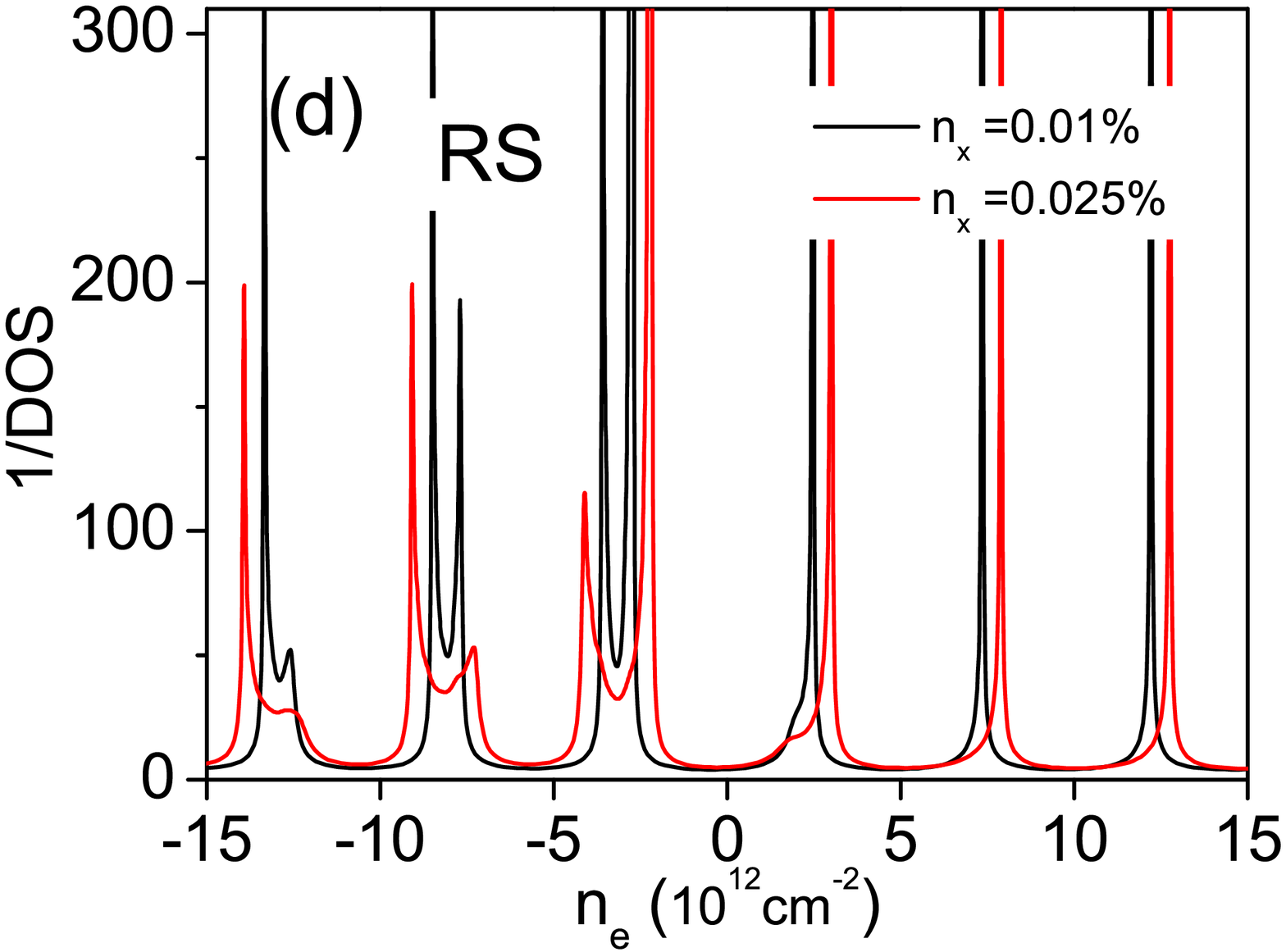}
  }
\mbox{
 \includegraphics[width=6cm]{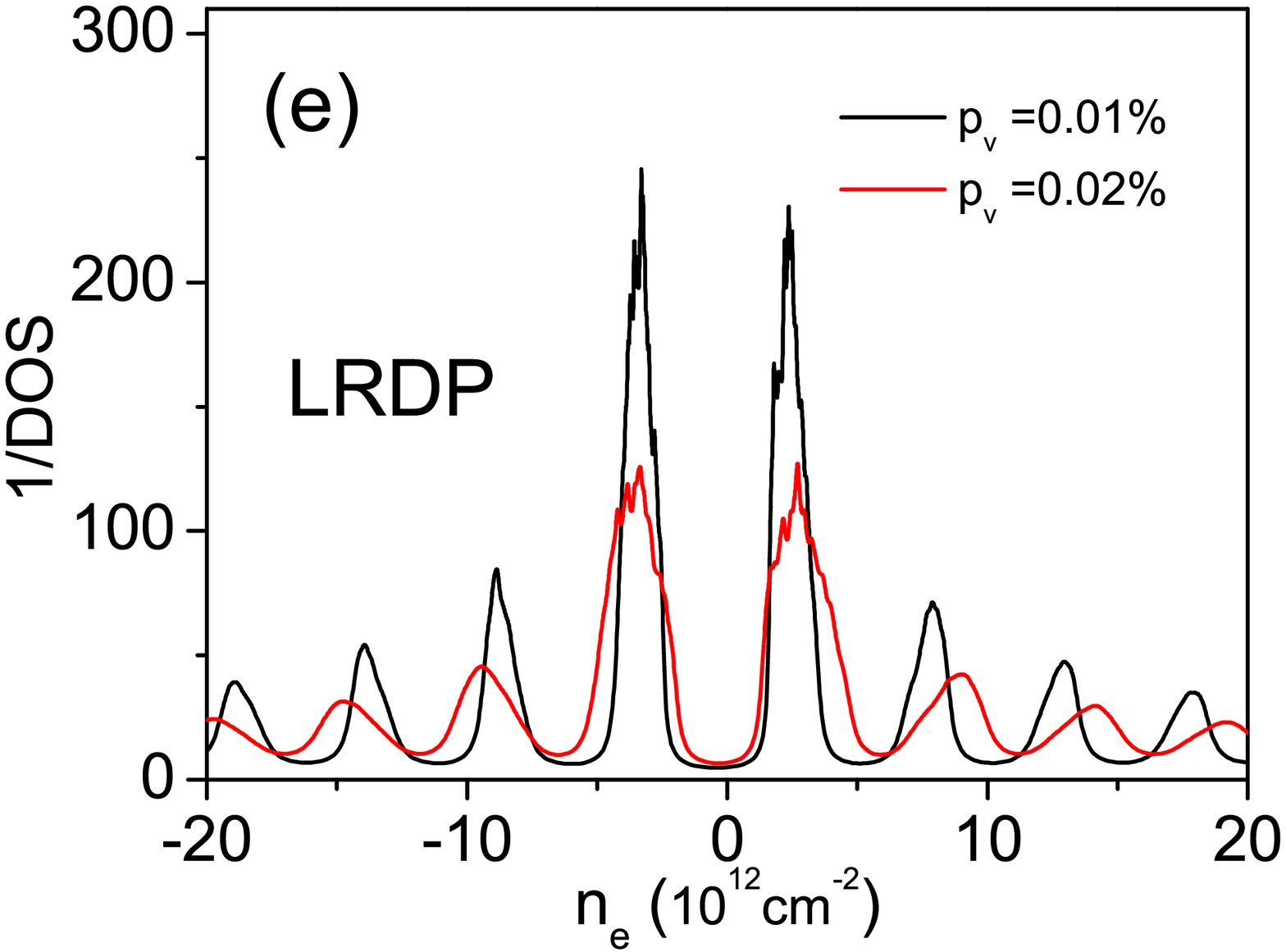}
 \includegraphics[width=6cm]{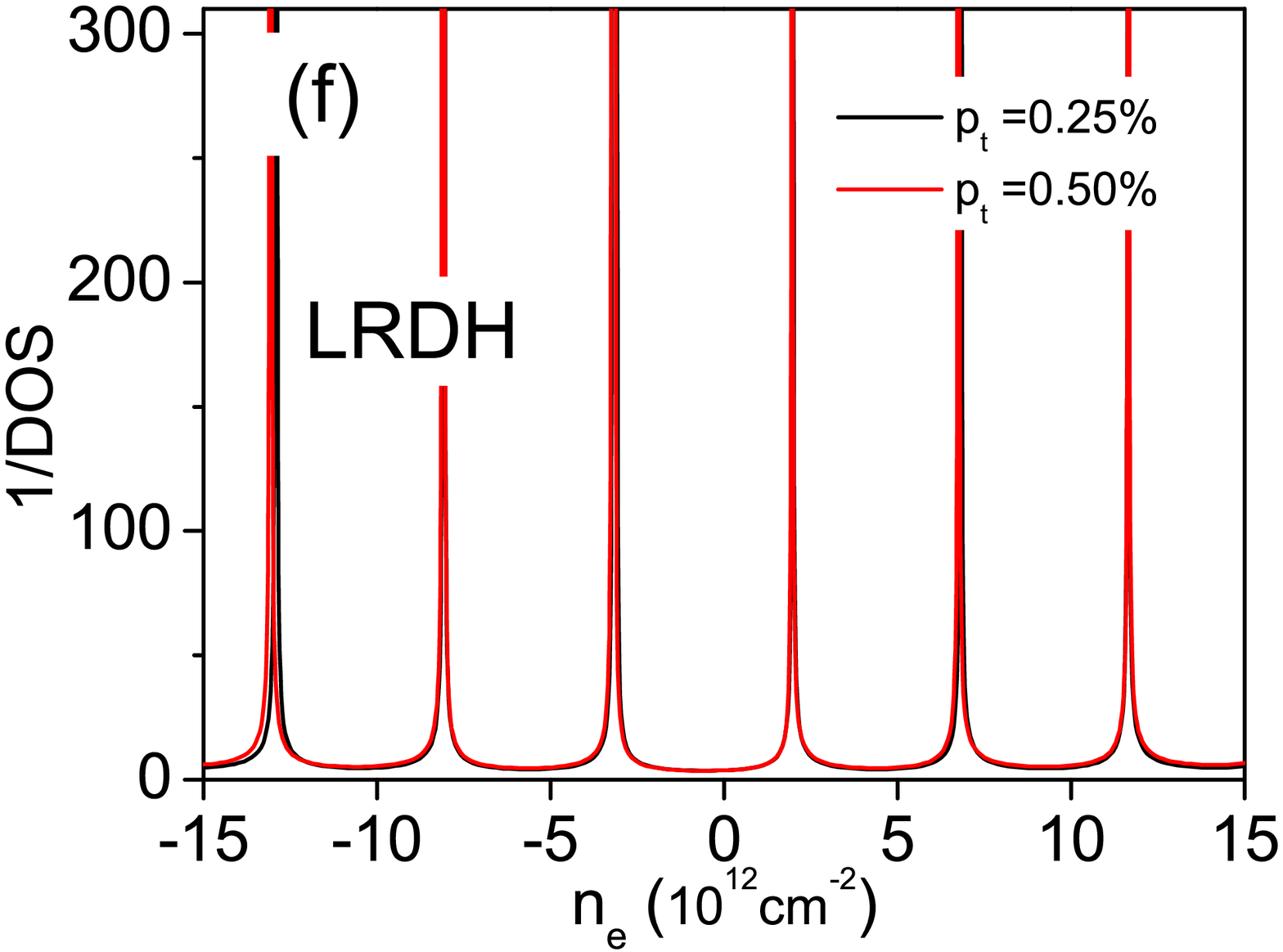}
  }
\caption{ (Color online) The reciprocal of DOS as a function of carry
density $n_e$ for disordered graphene in the presence of a uniform
perpendicular magnetic field ($B=50$T).}
\label{Fig:DOS2}
\end{figure*}
%________________________________________________

%________________________________________________
\begin{figure}[tb]
\includegraphics[width=0.8\columnwidth]{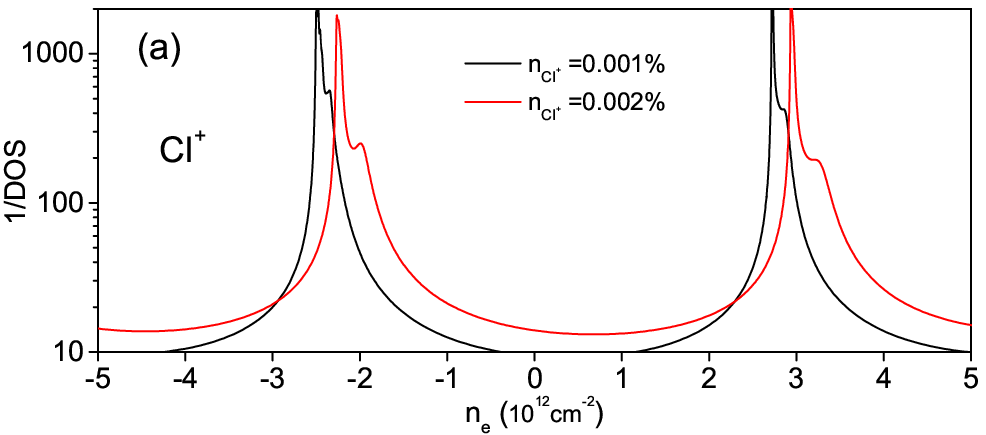} %
\includegraphics[width=0.8\columnwidth]{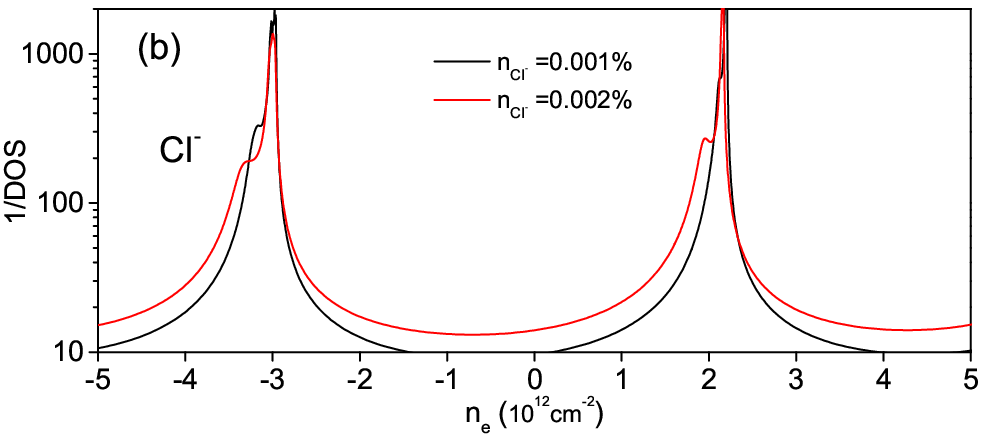}
\caption{ (Color online) The reciprocal of DOS as a function of carry
density $n_e$ for disordered graphene in the presence of a uniform
perpendicular magnetic field ($B=50$T).}
\label{Fig:DOS3}
\end{figure}
%________________________________________________

%\emph{Landau level spectrum }---

\section{LANDAU LEVEL SPECTRUM}

Finally we consider the electronic properties of graphene under a
perpendicular magnetic field ($B=50$T). The Landau quantization of the
energy levels leads to separated peaks, as shown in Fig.~\ref{Fig:DOS}. In
the presence of disorder, the peak amplitudes of the Landau levels (LL) are
reduced and the peaks become broader, except for LRDH in which the influence
of disorder is much weaker than for other types of disorders. The peak
profiles depend on the different sources of disorder. In general, for long
range disorder, the peak is still symmetric along its center, but for RS,
the changes are mainly restricted on the side with higher energy.
Furthermore, the LL spectrum exhibits electron-hole symmetry for CI$^{0}$
and LRDP, but becomes asymmetric for CI$^{+}$, CI$^{-}$ and RS. Especially,
there are two small peaks around the first Landau level on the hole side
shown in Fig.~\ref{Fig:DOS}(d), which has the same origin as for the zero LL
peaks, induced by RS~\cite{YRK10}. The differences that appear in the LL
spectrum also appear in quantum capacitance measurement, as the inverse of
the latter is proportional to DOS~\cite{QCAPA07,QCAPA09,QCAPA10,QCAPA14}.
Therefore, we also expect a huge effect of RS on the asymmetric quantum Hall
conductivity, a topic for future research.

The quantum capacitance $C_{q}$, which is defined as $C_{q}=\rho e^{2}$, can
be extracted experimentally from the total capacitance $C$ and the geometry
capacitance $C_{g}$ via $1/C_{q}=1/C-1/C_{g}$. In Fig.~\ref{Fig:DOS2} we
show the carrier dependence of $1/\rho $, which is proportional to $1/C_{q}$%
, for different types of disorders under the same magnetic field ($B=50$T).
Due to the presence of disorder, the peak amplitudes decrease significantly
except for the LRDH, in which the influence of random hopping is negligible.
The change of the spectrum profile for each type of disorder has similar
feature deduced from the corresponding DOS. Furthermore, some characters
become even more clear in the spectrum of $1/\rho $. For example, the
electron-hole asymmetry appeared in the presence of single-type charge
impurities (CI$^{+}$ or CI$^{-}$) is very special: the slopes of the peaks
on the hole and electron sides point to the same direction, depending on the
sign of CI (see a zoom of the first two peaks in Fig.~\ref{Fig:DOS3}). This
unique feature has also been observed in the experiments.\footnote{%
private communication with Konstantin Novoselov.}
\begin{table*}[t]
\centering
\makebox[1.0\linewidth]{
    \begin{tabular}{  C{0.04\linewidth} | C{0.07\linewidth}   | C{0.87\linewidth}}
    \hline\hline
    Ref.  & disorder   & Fingerprints\tabularnewline \hline
    [\onlinecite{Nuno14}] & LRDP & Symmetrical $\sigma \left( n_{e}\right) $ in Fig.2 (a), the minimum conductivity plateau in Fig. 2 (b), and the relation of mobility $\mu$ versus $n^*$ in Fig. 2 (c).    \tabularnewline \hline
    [\onlinecite{NG10}] &  RS &  Asymmetrical $\sigma \left( n_{e}\right) $ of the blue and red curves in Fig. 2 (a).     \tabularnewline \hline
    [\onlinecite{LB08}]   &   LRDP & A plateau in the doped optical spectroscopy in Fig. 2 (b), together with the corresponding relatively high mobility.   \tabularnewline \hline
    [\onlinecite{ChenCF2011}] &  CI$^{+}$  & A narrow plateau in the doped optical spectroscopy, together with a shift of the minimum conductivity to the electron side in Fig. 1.   \tabularnewline \hline
    [\onlinecite{Tan2007}]  & CI$^{+}$ &   The electron mobility is smaller than the hole one in Fig. 2 for samples K130, K145, K151; The minimum conductivity shifts to the electron side in Fig. 3.  \tabularnewline \hline
    [\onlinecite{ChenJH2008}] & CI$^{-}$   &  The hole conductivity is smaller than electron one and the minimum conductivity shifts to the hole side in Fig. 2.  \tabularnewline \hline
    [\onlinecite{Bolotin2008}]   & CI$^{+}$ &  The hole conductivity is larger than electron one in Fig. 1 (the sample before annealing).   \tabularnewline \hline
    [\onlinecite{Ren2012terahertz}]  & CI &  A narrow plateau in the doped optical spectroscopy in Fig. 3 (b).   \tabularnewline \hline
    \hline\hline
    \end{tabular}
    }
\caption{List of the dominant disorder source in different experimental
samples, identified by using the fingerprints appeared in the transport or
optical properties. The figures indicated in the table are these in the
corresponding reference.}
\label{table}
\end{table*}

%\textit{Conclusion---}

\section{DISCUSSION AND CONCLUSION}

%{\color{blue} % \textit{Conclusion---}
We have studied the effects of different types of disorders on the
electronic, transport and optical properties of graphene. By comparing the
results with and without the NNN hopping, we find that the NNN hopping has
negligible effect in combination with long-range disorder such as CI, LRDP
and LRDH, but that it changes the physical properties dramatically if RS are
present. In the latter case, we find that 1) there is an extra conductivity
plateau on the hole side, with a value larger than the minimum conductivity
at the neutrality point; 2) the carrier-density-dependent mobility does not
always drop with larger carrier density but instead, it reaches a minimum at
the edge of the conductivity plateau.
% in concert with the experimental observations~\cite{Tan2007};
3) a strong electron-hole asymmetry appears in the carrier-density-dependent
transport properties and Landau level spectrum; 4) the minimum conductivity
at the shifted Dirac point is no longer a constant, but drops to $4e^{2}/\pi
h$ when the impurity concentration is larger than $0.05\%$. For long-range
disorder, the minimum conductivity for $t^{\prime }=t/10$ is of the order of
$4e^{2}/h$ and increases with larger disorder concentration for CI and LRDP,
but remains the same for LRDP. The mobility always becomes smaller with
larger concentration of disorder, however, the minimum conductivity does not
follow the same rule, consistent with the transport measurement~\cite%
{Tan2007,Geim2007}. For doped graphene, the presence of disorder introduces
extra excitations below $2\mu _{F}$ but the profiles of the optical spectra
are different for different types of disorders.

As an example of using the fingerprints discussed in the main text, we
collect the dominant source of disorder in several well-known experiments
and list them in Table I. Different types of disorders such as CI (including
CI$^{0}$, CI$^{+}$ and CI$^{-}$), LRDP and RS have been identified in
different experiments, except for the LRDH which has been proved to have
negligible influence to the electronic properties. The results obtained in
Table I also suggest the dominant source of disorder may vary from sample
to sample.

In summary, we suggest that the different but characteristic features that
appear in the calculated electronic, transport and optical properties can be
used as fingerprints to identify the dominant sources of disorder in
graphene.

%\textit{Acknowledgments---}

\section{ACKNOWLEDGMENTS}

We thank the European Union Seventh Framework Programme under grant
agreement n604391 Graphene Flagship. The support by the China Scholarship
Council (CSC) and by the Stichting Fundamenteel Onderzoek der Materie (FOM)
and the Netherlands National Computing Facilities foundation (NCF) are
acknowledged. S.Y. and M.I.K. thank financial support from the European
Research Council Advanced Grant program (contract 338957).

%\bibliography{Bib_Graphene}
\bibliographystyle{apsrev}
\bibliography{Disorder_graphene}

\begin{thebibliography}{49}
\expandafter\ifx\csname natexlab\endcsname\relax\def\natexlab#1{#1}\fi
\expandafter\ifx\csname bibnamefont\endcsname\relax
  \def\bibnamefont#1{#1}\fi
\expandafter\ifx\csname bibfnamefont\endcsname\relax
  \def\bibfnamefont#1{#1}\fi
\expandafter\ifx\csname citenamefont\endcsname\relax
  \def\citenamefont#1{#1}\fi
\expandafter\ifx\csname url\endcsname\relax
  \def\url#1{\texttt{#1}}\fi
\expandafter\ifx\csname urlprefix\endcsname\relax\def\urlprefix{URL }\fi
\providecommand{\bibinfo}[2]{#2}
\providecommand{\eprint}[2][]{\url{#2}}

\bibitem[{\citenamefont{Peres}(2010)}]{Peres2010RMP}
\bibinfo{author}{\bibfnamefont{N.~M.~R.} \bibnamefont{Peres}},
  \bibinfo{journal}{Rev. Mod. Phys.} \textbf{\bibinfo{volume}{82}},
  \bibinfo{pages}{2673} (\bibinfo{year}{2010}).

\bibitem[{\citenamefont{Katsnelson}(2012)}]{KatsnelsonBook}
\bibinfo{author}{\bibfnamefont{M.~I.} \bibnamefont{Katsnelson}},
  \emph{\bibinfo{title}{Graphene: Carbon in Two Dimensions}}
  (\bibinfo{publisher}{Cambridge University Press}, \bibinfo{year}{2012}).

\bibitem[{\citenamefont{Ponomarenko et~al.}(2009)\citenamefont{Ponomarenko,
  Yang, Mohiuddin, Katsnelson, Novoselov, Morozov, Zhukov, Schedin, Hill, and
  Geim}}]{Ponomarenko2009}
\bibinfo{author}{\bibfnamefont{L.~A.} \bibnamefont{Ponomarenko}},
  \bibinfo{author}{\bibfnamefont{R.}~\bibnamefont{Yang}},
  \bibinfo{author}{\bibfnamefont{T.~M.} \bibnamefont{Mohiuddin}},
  \bibinfo{author}{\bibfnamefont{M.~I.} \bibnamefont{Katsnelson}},
  \bibinfo{author}{\bibfnamefont{K.~S.} \bibnamefont{Novoselov}},
  \bibinfo{author}{\bibfnamefont{S.~V.} \bibnamefont{Morozov}},
  \bibinfo{author}{\bibfnamefont{A.~A.} \bibnamefont{Zhukov}},
  \bibinfo{author}{\bibfnamefont{F.}~\bibnamefont{Schedin}},
  \bibinfo{author}{\bibfnamefont{E.~W.} \bibnamefont{Hill}}, \bibnamefont{and}
  \bibinfo{author}{\bibfnamefont{A.~K.} \bibnamefont{Geim}},
  \bibinfo{journal}{Phys. Rev. Lett.} \textbf{\bibinfo{volume}{102}},
  \bibinfo{pages}{206603} (\bibinfo{year}{2009}).

\bibitem[{\citenamefont{Couto et~al.}(2011)\citenamefont{Couto, Sac\'ep\'e, and
  Morpurgo}}]{Couto2011}
\bibinfo{author}{\bibfnamefont{N.~J.~G.} \bibnamefont{Couto}},
  \bibinfo{author}{\bibfnamefont{B.}~\bibnamefont{Sac\'ep\'e}},
  \bibnamefont{and} \bibinfo{author}{\bibfnamefont{A.~F.}
  \bibnamefont{Morpurgo}}, \bibinfo{journal}{Phys. Rev. Lett.}
  \textbf{\bibinfo{volume}{107}}, \bibinfo{pages}{225501}
  (\bibinfo{year}{2011}).

\bibitem[{\citenamefont{Katsnelson and Geim}(2008)}]{ripple2008}
\bibinfo{author}{\bibfnamefont{M.}~\bibnamefont{Katsnelson}} \bibnamefont{and}
  \bibinfo{author}{\bibfnamefont{A.}~\bibnamefont{Geim}},
  \bibinfo{journal}{Phil. Trans. R. Soc. A} \textbf{\bibinfo{volume}{366}},
  \bibinfo{pages}{195} (\bibinfo{year}{2008}).

\bibitem[{\citenamefont{Gibertini et~al.}(2010)\citenamefont{Gibertini,
  Tomadin, Polini, Fasolino, and Katsnelson}}]{Gibertini2010}
\bibinfo{author}{\bibfnamefont{M.}~\bibnamefont{Gibertini}},
  \bibinfo{author}{\bibfnamefont{A.}~\bibnamefont{Tomadin}},
  \bibinfo{author}{\bibfnamefont{M.}~\bibnamefont{Polini}},
  \bibinfo{author}{\bibfnamefont{A.}~\bibnamefont{Fasolino}}, \bibnamefont{and}
  \bibinfo{author}{\bibfnamefont{M.~I.} \bibnamefont{Katsnelson}},
  \bibinfo{journal}{Phys. Rev. B} \textbf{\bibinfo{volume}{81}},
  \bibinfo{pages}{125437} (\bibinfo{year}{2010}).

\bibitem[{\citenamefont{Gibertini et~al.}(2012)\citenamefont{Gibertini,
  Tomadin, Guinea, Katsnelson, and Polini}}]{Gibertini2012}
\bibinfo{author}{\bibfnamefont{M.}~\bibnamefont{Gibertini}},
  \bibinfo{author}{\bibfnamefont{A.}~\bibnamefont{Tomadin}},
  \bibinfo{author}{\bibfnamefont{F.}~\bibnamefont{Guinea}},
  \bibinfo{author}{\bibfnamefont{M.~I.} \bibnamefont{Katsnelson}},
  \bibnamefont{and} \bibinfo{author}{\bibfnamefont{M.}~\bibnamefont{Polini}},
  \bibinfo{journal}{Phys. Rev. B} \textbf{\bibinfo{volume}{85}},
  \bibinfo{pages}{201405} (\bibinfo{year}{2012}).

\bibitem[{\citenamefont{Couto et~al.}(2014)\citenamefont{Couto, Costanzo,
  Engels, Ki, Watanabe, Taniguchi, Stampfer, Guinea, and Morpurgo}}]{Nuno14}
\bibinfo{author}{\bibfnamefont{N.~J.~G.} \bibnamefont{Couto}},
  \bibinfo{author}{\bibfnamefont{D.}~\bibnamefont{Costanzo}},
  \bibinfo{author}{\bibfnamefont{S.}~\bibnamefont{Engels}},
  \bibinfo{author}{\bibfnamefont{D.-K.} \bibnamefont{Ki}},
  \bibinfo{author}{\bibfnamefont{K.}~\bibnamefont{Watanabe}},
  \bibinfo{author}{\bibfnamefont{T.}~\bibnamefont{Taniguchi}},
  \bibinfo{author}{\bibfnamefont{C.}~\bibnamefont{Stampfer}},
  \bibinfo{author}{\bibfnamefont{F.}~\bibnamefont{Guinea}}, \bibnamefont{and}
  \bibinfo{author}{\bibfnamefont{A.~F.} \bibnamefont{Morpurgo}},
  \bibinfo{journal}{Phys. Rev. X} \textbf{\bibinfo{volume}{4}},
  \bibinfo{pages}{041019} (\bibinfo{year}{2014}).

\bibitem[{\citenamefont{Vozmediano et~al.}(2010)\citenamefont{Vozmediano,
  Katsnelson, and Guinea}}]{Vozmediano2010}
\bibinfo{author}{\bibfnamefont{M.~A.~H.} \bibnamefont{Vozmediano}},
  \bibinfo{author}{\bibfnamefont{M.~I.} \bibnamefont{Katsnelson}},
  \bibnamefont{and} \bibinfo{author}{\bibfnamefont{F.}~\bibnamefont{Guinea}},
  \bibinfo{journal}{Physics Reports} \textbf{\bibinfo{volume}{496}},
  \bibinfo{pages}{109 } (\bibinfo{year}{2010}).

\bibitem[{\citenamefont{Ni et~al.}(2010)\citenamefont{Ni, Ponomarenko, Nair,
  Yang, Anissimova, Grigorieva, Schedin, Blake, Shen, Hill et~al.}}]{NG10}
\bibinfo{author}{\bibfnamefont{Z.}~\bibnamefont{Ni}},
  \bibinfo{author}{\bibfnamefont{L.}~\bibnamefont{Ponomarenko}},
  \bibinfo{author}{\bibfnamefont{R.}~\bibnamefont{Nair}},
  \bibinfo{author}{\bibfnamefont{R.}~\bibnamefont{Yang}},
  \bibinfo{author}{\bibfnamefont{S.}~\bibnamefont{Anissimova}},
  \bibinfo{author}{\bibfnamefont{I.}~\bibnamefont{Grigorieva}},
  \bibinfo{author}{\bibfnamefont{F.}~\bibnamefont{Schedin}},
  \bibinfo{author}{\bibfnamefont{P.}~\bibnamefont{Blake}},
  \bibinfo{author}{\bibfnamefont{Z.}~\bibnamefont{Shen}},
  \bibinfo{author}{\bibfnamefont{E.}~\bibnamefont{Hill}}, \bibnamefont{et~al.},
  \bibinfo{journal}{Nano letters} \textbf{\bibinfo{volume}{10}},
  \bibinfo{pages}{3868} (\bibinfo{year}{2010}).

\bibitem[{\citenamefont{Wehling et~al.}(2010)\citenamefont{Wehling, Yuan,
  Lichtenstein, Geim, and Katsnelson}}]{WK10}
\bibinfo{author}{\bibfnamefont{T.~O.} \bibnamefont{Wehling}},
  \bibinfo{author}{\bibfnamefont{S.}~\bibnamefont{Yuan}},
  \bibinfo{author}{\bibfnamefont{A.~I.} \bibnamefont{Lichtenstein}},
  \bibinfo{author}{\bibfnamefont{A.~K.} \bibnamefont{Geim}}, \bibnamefont{and}
  \bibinfo{author}{\bibfnamefont{M.~I.} \bibnamefont{Katsnelson}},
  \bibinfo{journal}{Phys. Rev. Lett.} \textbf{\bibinfo{volume}{105}},
  \bibinfo{pages}{056802} (\bibinfo{year}{2010}).

\bibitem[{\citenamefont{Orlita and Potemski}(2010)}]{OP10}
\bibinfo{author}{\bibfnamefont{M.}~\bibnamefont{Orlita}} \bibnamefont{and}
  \bibinfo{author}{\bibfnamefont{M.}~\bibnamefont{Potemski}},
  \bibinfo{journal}{Semiconductor Science and Technology}
  \textbf{\bibinfo{volume}{25}}, \bibinfo{pages}{063001}
  (\bibinfo{year}{2010}).

\bibitem[{\citenamefont{Wang et~al.}(2008)\citenamefont{Wang, Zhang, Tian,
  Girit, Zettl, Crommie, and Shen}}]{WangF2008}
\bibinfo{author}{\bibfnamefont{F.}~\bibnamefont{Wang}},
  \bibinfo{author}{\bibfnamefont{Y.}~\bibnamefont{Zhang}},
  \bibinfo{author}{\bibfnamefont{C.}~\bibnamefont{Tian}},
  \bibinfo{author}{\bibfnamefont{C.}~\bibnamefont{Girit}},
  \bibinfo{author}{\bibfnamefont{A.}~\bibnamefont{Zettl}},
  \bibinfo{author}{\bibfnamefont{M.}~\bibnamefont{Crommie}}, \bibnamefont{and}
  \bibinfo{author}{\bibfnamefont{Y.~R.} \bibnamefont{Shen}},
  \bibinfo{journal}{Science} \textbf{\bibinfo{volume}{320}},
  \bibinfo{pages}{206} (\bibinfo{year}{2008}).

\bibitem[{\citenamefont{Li et~al.}(2008)\citenamefont{Li, Henriksen, Jiang,
  Hao, Martin, Kim, Stormer, and Basov}}]{LB08}
\bibinfo{author}{\bibfnamefont{Z.}~\bibnamefont{Li}},
  \bibinfo{author}{\bibfnamefont{E.~A.} \bibnamefont{Henriksen}},
  \bibinfo{author}{\bibfnamefont{Z.}~\bibnamefont{Jiang}},
  \bibinfo{author}{\bibfnamefont{Z.}~\bibnamefont{Hao}},
  \bibinfo{author}{\bibfnamefont{M.~C.} \bibnamefont{Martin}},
  \bibinfo{author}{\bibfnamefont{P.}~\bibnamefont{Kim}},
  \bibinfo{author}{\bibfnamefont{H.}~\bibnamefont{Stormer}}, \bibnamefont{and}
  \bibinfo{author}{\bibfnamefont{D.~N.} \bibnamefont{Basov}},
  \bibinfo{journal}{Nature Physics} \textbf{\bibinfo{volume}{4}},
  \bibinfo{pages}{532} (\bibinfo{year}{2008}).

\bibitem[{\citenamefont{Chen et~al.}(2011)\citenamefont{Chen, Park, Boudouris,
  Horng, Geng, Girit, Zettl, Crommie, Segalman, Louie et~al.}}]{ChenCF2011}
\bibinfo{author}{\bibfnamefont{C.-F.} \bibnamefont{Chen}},
  \bibinfo{author}{\bibfnamefont{C.-H.} \bibnamefont{Park}},
  \bibinfo{author}{\bibfnamefont{B.~W.} \bibnamefont{Boudouris}},
  \bibinfo{author}{\bibfnamefont{J.}~\bibnamefont{Horng}},
  \bibinfo{author}{\bibfnamefont{B.}~\bibnamefont{Geng}},
  \bibinfo{author}{\bibfnamefont{C.}~\bibnamefont{Girit}},
  \bibinfo{author}{\bibfnamefont{A.}~\bibnamefont{Zettl}},
  \bibinfo{author}{\bibfnamefont{M.~F.} \bibnamefont{Crommie}},
  \bibinfo{author}{\bibfnamefont{R.~A.} \bibnamefont{Segalman}},
  \bibinfo{author}{\bibfnamefont{S.~G.} \bibnamefont{Louie}},
  \bibnamefont{et~al.}, \bibinfo{journal}{Nature}
  \textbf{\bibinfo{volume}{471}}, \bibinfo{pages}{617} (\bibinfo{year}{2011}).

\bibitem[{\citenamefont{Ando et~al.}(2002)\citenamefont{Ando, Zheng, and
  Suzuura}}]{Ando2002}
\bibinfo{author}{\bibfnamefont{T.}~\bibnamefont{Ando}},
  \bibinfo{author}{\bibfnamefont{Y.}~\bibnamefont{Zheng}}, \bibnamefont{and}
  \bibinfo{author}{\bibfnamefont{H.}~\bibnamefont{Suzuura}},
  \bibinfo{journal}{Journal of the Physical Society of Japan}
  \textbf{\bibinfo{volume}{71}}, \bibinfo{pages}{1318} (\bibinfo{year}{2002}).

\bibitem[{\citenamefont{Gr\"uneis et~al.}(2003)\citenamefont{Gr\"uneis, Saito,
  Samsonidze, Kimura, Pimenta, Jorio, Filho, Dresselhaus, and
  Dresselhaus}}]{Gruneis2003}
\bibinfo{author}{\bibfnamefont{A.}~\bibnamefont{Gr\"uneis}},
  \bibinfo{author}{\bibfnamefont{R.}~\bibnamefont{Saito}},
  \bibinfo{author}{\bibfnamefont{G.~G.} \bibnamefont{Samsonidze}},
  \bibinfo{author}{\bibfnamefont{T.}~\bibnamefont{Kimura}},
  \bibinfo{author}{\bibfnamefont{M.~A.} \bibnamefont{Pimenta}},
  \bibinfo{author}{\bibfnamefont{A.}~\bibnamefont{Jorio}},
  \bibinfo{author}{\bibfnamefont{A.~G.~S.} \bibnamefont{Filho}},
  \bibinfo{author}{\bibfnamefont{G.}~\bibnamefont{Dresselhaus}},
  \bibnamefont{and} \bibinfo{author}{\bibfnamefont{M.~S.}
  \bibnamefont{Dresselhaus}}, \bibinfo{journal}{Phys. Rev. B}
  \textbf{\bibinfo{volume}{67}}, \bibinfo{pages}{165402}
  (\bibinfo{year}{2003}).

\bibitem[{\citenamefont{Peres et~al.}(2006)\citenamefont{Peres, Guinea, and
  Castro~Neto}}]{Peres2006}
\bibinfo{author}{\bibfnamefont{N.~M.~R.} \bibnamefont{Peres}},
  \bibinfo{author}{\bibfnamefont{F.}~\bibnamefont{Guinea}}, \bibnamefont{and}
  \bibinfo{author}{\bibfnamefont{A.~H.} \bibnamefont{Castro~Neto}},
  \bibinfo{journal}{Phys. Rev. B} \textbf{\bibinfo{volume}{73}},
  \bibinfo{pages}{125411} (\bibinfo{year}{2006}).

\bibitem[{\citenamefont{Gusynin et~al.}(2006)\citenamefont{Gusynin, Sharapov,
  and Carbotte}}]{Gusynin2006}
\bibinfo{author}{\bibfnamefont{V.~P.} \bibnamefont{Gusynin}},
  \bibinfo{author}{\bibfnamefont{S.~G.} \bibnamefont{Sharapov}},
  \bibnamefont{and} \bibinfo{author}{\bibfnamefont{J.~P.}
  \bibnamefont{Carbotte}}, \bibinfo{journal}{Phys. Rev. Lett.}
  \textbf{\bibinfo{volume}{96}}, \bibinfo{pages}{256802}
  (\bibinfo{year}{2006}).

\bibitem[{\citenamefont{Stauber et~al.}(2007)\citenamefont{Stauber, Peres, and
  Guinea}}]{SPG07}
\bibinfo{author}{\bibfnamefont{T.}~\bibnamefont{Stauber}},
  \bibinfo{author}{\bibfnamefont{N.~M.~R.} \bibnamefont{Peres}},
  \bibnamefont{and} \bibinfo{author}{\bibfnamefont{F.}~\bibnamefont{Guinea}},
  \bibinfo{journal}{Phys. Rev. B} \textbf{\bibinfo{volume}{76}},
  \bibinfo{pages}{205423} (\bibinfo{year}{2007}).

\bibitem[{\citenamefont{Gusynin et~al.}(2007)\citenamefont{Gusynin, Sharapov,
  and Carbotte}}]{Gusynin2007}
\bibinfo{author}{\bibfnamefont{V.~P.} \bibnamefont{Gusynin}},
  \bibinfo{author}{\bibfnamefont{S.~G.} \bibnamefont{Sharapov}},
  \bibnamefont{and} \bibinfo{author}{\bibfnamefont{J.~P.}
  \bibnamefont{Carbotte}}, \bibinfo{journal}{International Journal of Modern
  Physics B} \textbf{\bibinfo{volume}{21}}, \bibinfo{pages}{4611}
  (\bibinfo{year}{2007}).

\bibitem[{\citenamefont{Stauber
  et~al.}(2008{\natexlab{a}})\citenamefont{Stauber, Peres, and
  Geim}}]{Stauber2008}
\bibinfo{author}{\bibfnamefont{T.}~\bibnamefont{Stauber}},
  \bibinfo{author}{\bibfnamefont{N.~M.~R.} \bibnamefont{Peres}},
  \bibnamefont{and} \bibinfo{author}{\bibfnamefont{A.~K.} \bibnamefont{Geim}},
  \bibinfo{journal}{Phys. Rev. B} \textbf{\bibinfo{volume}{78}},
  \bibinfo{pages}{085432} (\bibinfo{year}{2008}{\natexlab{a}}).

\bibitem[{\citenamefont{Stauber
  et~al.}(2008{\natexlab{b}})\citenamefont{Stauber, Peres, and
  Castro~Neto}}]{Stauber2008b}
\bibinfo{author}{\bibfnamefont{T.}~\bibnamefont{Stauber}},
  \bibinfo{author}{\bibfnamefont{N.~M.~R.} \bibnamefont{Peres}},
  \bibnamefont{and} \bibinfo{author}{\bibfnamefont{A.~H.}
  \bibnamefont{Castro~Neto}}, \bibinfo{journal}{Phys. Rev. B}
  \textbf{\bibinfo{volume}{78}}, \bibinfo{pages}{085418}
  (\bibinfo{year}{2008}{\natexlab{b}}).

\bibitem[{\citenamefont{Min and MacDonald}(2009)}]{MinHK2009}
\bibinfo{author}{\bibfnamefont{H.}~\bibnamefont{Min}} \bibnamefont{and}
  \bibinfo{author}{\bibfnamefont{A.~H.} \bibnamefont{MacDonald}},
  \bibinfo{journal}{Phys. Rev. Lett.} \textbf{\bibinfo{volume}{103}},
  \bibinfo{pages}{067402} (\bibinfo{year}{2009}).

\bibitem[{\citenamefont{Mak et~al.}(2008)\citenamefont{Mak, Sfeir, Wu, Lui,
  Misewich, and Heinz}}]{Mak2008}
\bibinfo{author}{\bibfnamefont{K.~F.} \bibnamefont{Mak}},
  \bibinfo{author}{\bibfnamefont{M.~Y.} \bibnamefont{Sfeir}},
  \bibinfo{author}{\bibfnamefont{Y.}~\bibnamefont{Wu}},
  \bibinfo{author}{\bibfnamefont{C.~H.} \bibnamefont{Lui}},
  \bibinfo{author}{\bibfnamefont{J.~A.} \bibnamefont{Misewich}},
  \bibnamefont{and} \bibinfo{author}{\bibfnamefont{T.~F.} \bibnamefont{Heinz}},
  \bibinfo{journal}{Phys. Rev. Lett.} \textbf{\bibinfo{volume}{101}},
  \bibinfo{pages}{196405} (\bibinfo{year}{2008}).

\bibitem[{\citenamefont{Yuan et~al.}(2011)\citenamefont{Yuan, Rold\'an,
  De~Raedt, and Katsnelson}}]{Yuan2011}
\bibinfo{author}{\bibfnamefont{S.}~\bibnamefont{Yuan}},
  \bibinfo{author}{\bibfnamefont{R.}~\bibnamefont{Rold\'an}},
  \bibinfo{author}{\bibfnamefont{H.}~\bibnamefont{De~Raedt}}, \bibnamefont{and}
  \bibinfo{author}{\bibfnamefont{M.~I.} \bibnamefont{Katsnelson}},
  \bibinfo{journal}{Phys. Rev. B} \textbf{\bibinfo{volume}{84}},
  \bibinfo{pages}{195418} (\bibinfo{year}{2011}).

\bibitem[{\citenamefont{Castro~Neto et~al.}(2009)\citenamefont{Castro~Neto,
  Guinea, Peres, Novoselov, and Geim}}]{CG09}
\bibinfo{author}{\bibfnamefont{A.~H.} \bibnamefont{Castro~Neto}},
  \bibinfo{author}{\bibfnamefont{F.}~\bibnamefont{Guinea}},
  \bibinfo{author}{\bibfnamefont{N.~M.~R.} \bibnamefont{Peres}},
  \bibinfo{author}{\bibfnamefont{K.~S.} \bibnamefont{Novoselov}},
  \bibnamefont{and} \bibinfo{author}{\bibfnamefont{A.~K.} \bibnamefont{Geim}},
  \bibinfo{journal}{Rev. Mod. Phys.} \textbf{\bibinfo{volume}{81}},
  \bibinfo{pages}{109} (\bibinfo{year}{2009}).

\bibitem[{\citenamefont{Kretinin et~al.}(2013)\citenamefont{Kretinin, Yu,
  Jalil, Cao, Withers, Mishchenko, Katsnelson, Novoselov, Geim, and
  Guinea}}]{capacitance2013}
\bibinfo{author}{\bibfnamefont{A.}~\bibnamefont{Kretinin}},
  \bibinfo{author}{\bibfnamefont{G.~L.} \bibnamefont{Yu}},
  \bibinfo{author}{\bibfnamefont{R.}~\bibnamefont{Jalil}},
  \bibinfo{author}{\bibfnamefont{Y.}~\bibnamefont{Cao}},
  \bibinfo{author}{\bibfnamefont{F.}~\bibnamefont{Withers}},
  \bibinfo{author}{\bibfnamefont{A.}~\bibnamefont{Mishchenko}},
  \bibinfo{author}{\bibfnamefont{M.~I.} \bibnamefont{Katsnelson}},
  \bibinfo{author}{\bibfnamefont{K.~S.} \bibnamefont{Novoselov}},
  \bibinfo{author}{\bibfnamefont{A.~K.} \bibnamefont{Geim}}, \bibnamefont{and}
  \bibinfo{author}{\bibfnamefont{F.}~\bibnamefont{Guinea}},
  \bibinfo{journal}{Phys. Rev. B} \textbf{\bibinfo{volume}{88}},
  \bibinfo{pages}{165427} (\bibinfo{year}{2013}).

\bibitem[{\citenamefont{Pereira et~al.}(2007)\citenamefont{Pereira, Nilsson,
  and Castro~Neto}}]{Pereira2007}
\bibinfo{author}{\bibfnamefont{V.~M.} \bibnamefont{Pereira}},
  \bibinfo{author}{\bibfnamefont{J.}~\bibnamefont{Nilsson}}, \bibnamefont{and}
  \bibinfo{author}{\bibfnamefont{A.~H.} \bibnamefont{Castro~Neto}},
  \bibinfo{journal}{Phys. Rev. Lett.} \textbf{\bibinfo{volume}{99}},
  \bibinfo{pages}{166802} (\bibinfo{year}{2007}).

\bibitem[{\citenamefont{Robinson et~al.}(2008)\citenamefont{Robinson,
  Schomerus, Oroszl\'any, and Fal'ko}}]{Robinson08}
\bibinfo{author}{\bibfnamefont{J.~P.} \bibnamefont{Robinson}},
  \bibinfo{author}{\bibfnamefont{H.}~\bibnamefont{Schomerus}},
  \bibinfo{author}{\bibfnamefont{L.}~\bibnamefont{Oroszl\'any}},
  \bibnamefont{and} \bibinfo{author}{\bibfnamefont{V.~I.}
  \bibnamefont{Fal'ko}}, \bibinfo{journal}{Phys. Rev. Lett.}
  \textbf{\bibinfo{volume}{101}}, \bibinfo{pages}{196803}
  (\bibinfo{year}{2008}).

\bibitem[{\citenamefont{Yuan et~al.}(2010)\citenamefont{Yuan, De~Raedt, and
  Katsnelson}}]{YRK10}
\bibinfo{author}{\bibfnamefont{S.}~\bibnamefont{Yuan}},
  \bibinfo{author}{\bibfnamefont{H.}~\bibnamefont{De~Raedt}}, \bibnamefont{and}
  \bibinfo{author}{\bibfnamefont{M.~I.} \bibnamefont{Katsnelson}},
  \bibinfo{journal}{Phys. Rev. B} \textbf{\bibinfo{volume}{82}},
  \bibinfo{pages}{115448} (\bibinfo{year}{2010}).

\bibitem[{\citenamefont{Hams and De~Raedt}(2000)}]{HR00}
\bibinfo{author}{\bibfnamefont{A.}~\bibnamefont{Hams}} \bibnamefont{and}
  \bibinfo{author}{\bibfnamefont{H.}~\bibnamefont{De~Raedt}},
  \bibinfo{journal}{Phys. Rev. E} \textbf{\bibinfo{volume}{62}},
  \bibinfo{pages}{4365} (\bibinfo{year}{2000}).

\bibitem[{\citenamefont{Yuan et~al.}(2012)\citenamefont{Yuan, Wehling,
  Lichtenstein, and Katsnelson}}]{Yuan2012}
\bibinfo{author}{\bibfnamefont{S.}~\bibnamefont{Yuan}},
  \bibinfo{author}{\bibfnamefont{T.~O.} \bibnamefont{Wehling}},
  \bibinfo{author}{\bibfnamefont{A.~I.} \bibnamefont{Lichtenstein}},
  \bibnamefont{and} \bibinfo{author}{\bibfnamefont{M.~I.}
  \bibnamefont{Katsnelson}}, \bibinfo{journal}{Phys. Rev. Lett.}
  \textbf{\bibinfo{volume}{109}}, \bibinfo{pages}{156601}
  (\bibinfo{year}{2012}).

\bibitem[{\citenamefont{Cresti et~al.}(2013)\citenamefont{Cresti, Ortmann,
  Louvet, Van~Tuan, and Roche}}]{Cresti2013}
\bibinfo{author}{\bibfnamefont{A.}~\bibnamefont{Cresti}},
  \bibinfo{author}{\bibfnamefont{F.}~\bibnamefont{Ortmann}},
  \bibinfo{author}{\bibfnamefont{T.}~\bibnamefont{Louvet}},
  \bibinfo{author}{\bibfnamefont{D.}~\bibnamefont{Van~Tuan}}, \bibnamefont{and}
  \bibinfo{author}{\bibfnamefont{S.}~\bibnamefont{Roche}},
  \bibinfo{journal}{Phys. Rev. Lett.} \textbf{\bibinfo{volume}{110}},
  \bibinfo{pages}{196601} (\bibinfo{year}{2013}).

\bibitem[{\citenamefont{Trambly~de Laissardi\`ere and
  Mayou}(2013)}]{Laissa2013}
\bibinfo{author}{\bibfnamefont{G.}~\bibnamefont{Trambly~de Laissardi\`ere}}
  \bibnamefont{and} \bibinfo{author}{\bibfnamefont{D.}~\bibnamefont{Mayou}},
  \bibinfo{journal}{Phys. Rev. Lett.} \textbf{\bibinfo{volume}{111}},
  \bibinfo{pages}{146601} (\bibinfo{year}{2013}).

\bibitem[{\citenamefont{Trambly~de Laissardi\`ere and
  Mayou}(2014)}]{Laissa2014}
\bibinfo{author}{\bibfnamefont{G.}~\bibnamefont{Trambly~de Laissardi\`ere}}
  \bibnamefont{and} \bibinfo{author}{\bibfnamefont{D.}~\bibnamefont{Mayou}},
  \bibinfo{journal}{Advances in Natural Sciences: Nanoscience and
  Nanotechnology} \textbf{\bibinfo{volume}{5}}, \bibinfo{pages}{015007}
  (\bibinfo{year}{2014}).

\bibitem[{\citenamefont{Pereira et~al.}(2008)\citenamefont{Pereira, Lopes~dos
  Santos, and Castro~Neto}}]{Pereira2008}
\bibinfo{author}{\bibfnamefont{V.~M.} \bibnamefont{Pereira}},
  \bibinfo{author}{\bibfnamefont{J.~M.~B.} \bibnamefont{Lopes~dos Santos}},
  \bibnamefont{and} \bibinfo{author}{\bibfnamefont{A.~H.}
  \bibnamefont{Castro~Neto}}, \bibinfo{journal}{Phys. Rev. B}
  \textbf{\bibinfo{volume}{77}}, \bibinfo{pages}{115109}
  (\bibinfo{year}{2008}).

\bibitem[{\citenamefont{Zhu and Lv}(2013)}]{ZhuW2013}
\bibinfo{author}{\bibfnamefont{W.}~\bibnamefont{Zhu}} \bibnamefont{and}
  \bibinfo{author}{\bibfnamefont{B.}~\bibnamefont{Lv}},
  \bibinfo{journal}{Physics Letters A} \textbf{\bibinfo{volume}{377}},
  \bibinfo{pages}{1649} (\bibinfo{year}{2013}).

\bibitem[{\citenamefont{Tan et~al.}(2007)\citenamefont{Tan, Zhang, Bolotin,
  Zhao, Adam, Hwang, Das~Sarma, Stormer, and Kim}}]{Tan2007}
\bibinfo{author}{\bibfnamefont{Y.-W.} \bibnamefont{Tan}},
  \bibinfo{author}{\bibfnamefont{Y.}~\bibnamefont{Zhang}},
  \bibinfo{author}{\bibfnamefont{K.}~\bibnamefont{Bolotin}},
  \bibinfo{author}{\bibfnamefont{Y.}~\bibnamefont{Zhao}},
  \bibinfo{author}{\bibfnamefont{S.}~\bibnamefont{Adam}},
  \bibinfo{author}{\bibfnamefont{E.~H.} \bibnamefont{Hwang}},
  \bibinfo{author}{\bibfnamefont{S.}~\bibnamefont{Das~Sarma}},
  \bibinfo{author}{\bibfnamefont{H.~L.} \bibnamefont{Stormer}},
  \bibnamefont{and} \bibinfo{author}{\bibfnamefont{P.}~\bibnamefont{Kim}},
  \bibinfo{journal}{Phys. Rev. Lett.} \textbf{\bibinfo{volume}{99}},
  \bibinfo{pages}{246803} (\bibinfo{year}{2007}).

\bibitem[{\citenamefont{Chen et~al.}(2008)\citenamefont{Chen, Jang, Adam,
  Fuhrer, Williams, and Ishigami}}]{ChenJH2008}
\bibinfo{author}{\bibfnamefont{J.-H.} \bibnamefont{Chen}},
  \bibinfo{author}{\bibfnamefont{C.}~\bibnamefont{Jang}},
  \bibinfo{author}{\bibfnamefont{S.}~\bibnamefont{Adam}},
  \bibinfo{author}{\bibfnamefont{M.}~\bibnamefont{Fuhrer}},
  \bibinfo{author}{\bibfnamefont{E.}~\bibnamefont{Williams}}, \bibnamefont{and}
  \bibinfo{author}{\bibfnamefont{M.}~\bibnamefont{Ishigami}},
  \bibinfo{journal}{Nature Physics} \textbf{\bibinfo{volume}{4}},
  \bibinfo{pages}{377} (\bibinfo{year}{2008}).

\bibitem[{\citenamefont{Geim and Novoselov}(2007)}]{Geim2007}
\bibinfo{author}{\bibfnamefont{A.~K.} \bibnamefont{Geim}} \bibnamefont{and}
  \bibinfo{author}{\bibfnamefont{K.~S.} \bibnamefont{Novoselov}},
  \bibinfo{journal}{Nature materials} \textbf{\bibinfo{volume}{6}},
  \bibinfo{pages}{183} (\bibinfo{year}{2007}).

\bibitem[{\citenamefont{Evers and Mirlin}(2008)}]{Andersontransitions2008}
\bibinfo{author}{\bibfnamefont{F.}~\bibnamefont{Evers}} \bibnamefont{and}
  \bibinfo{author}{\bibfnamefont{A.~D.} \bibnamefont{Mirlin}},
  \bibinfo{journal}{Rev. Mod. Phys.} \textbf{\bibinfo{volume}{80}},
  \bibinfo{pages}{1355} (\bibinfo{year}{2008}).

\bibitem[{\citenamefont{Ishihara}(1971)}]{Ishihara1971}
\bibinfo{author}{\bibfnamefont{A.}~\bibnamefont{Ishihara}},
  \emph{\bibinfo{title}{Statistical Physics}} (\bibinfo{publisher}{Academic
  Press, New York}, \bibinfo{year}{1971}).

\bibitem[{\citenamefont{Fang et~al.}(2007)\citenamefont{Fang, Konar, Xing, and
  Jena}}]{QCAPA07}
\bibinfo{author}{\bibfnamefont{T.}~\bibnamefont{Fang}},
  \bibinfo{author}{\bibfnamefont{A.}~\bibnamefont{Konar}},
  \bibinfo{author}{\bibfnamefont{H.}~\bibnamefont{Xing}}, \bibnamefont{and}
  \bibinfo{author}{\bibfnamefont{D.}~\bibnamefont{Jena}},
  \bibinfo{journal}{Applied Physics Letters} \textbf{\bibinfo{volume}{91}},
  \bibinfo{pages}{092109} (\bibinfo{year}{2007}).

\bibitem[{\citenamefont{Xia et~al.}(2009)\citenamefont{Xia, Chen, Li, and
  Tao}}]{QCAPA09}
\bibinfo{author}{\bibfnamefont{J.}~\bibnamefont{Xia}},
  \bibinfo{author}{\bibfnamefont{F.}~\bibnamefont{Chen}},
  \bibinfo{author}{\bibfnamefont{J.}~\bibnamefont{Li}}, \bibnamefont{and}
  \bibinfo{author}{\bibfnamefont{N.}~\bibnamefont{Tao}},
  \bibinfo{journal}{Nature Nanotech.} \textbf{\bibinfo{volume}{4}},
  \bibinfo{pages}{505} (\bibinfo{year}{2009}).

\bibitem[{\citenamefont{Dr\"oscherscher
  et~al.}(2010)\citenamefont{Dr\"oscherscher, Roulleau, Molitor, Studerus,
  Stampfer, Ensslin, and Ihn}}]{QCAPA10}
\bibinfo{author}{\bibfnamefont{S.}~\bibnamefont{Dr\"oscherscher}},
  \bibinfo{author}{\bibfnamefont{P.}~\bibnamefont{Roulleau}},
  \bibinfo{author}{\bibfnamefont{F.}~\bibnamefont{Molitor}},
  \bibinfo{author}{\bibfnamefont{P.}~\bibnamefont{Studerus}},
  \bibinfo{author}{\bibfnamefont{C.}~\bibnamefont{Stampfer}},
  \bibinfo{author}{\bibfnamefont{K.}~\bibnamefont{Ensslin}}, \bibnamefont{and}
  \bibinfo{author}{\bibfnamefont{T.}~\bibnamefont{Ihn}},
  \bibinfo{journal}{Appl. Phys. Lett} \textbf{\bibinfo{volume}{96}},
  \bibinfo{pages}{152104} (\bibinfo{year}{2010}).

\bibitem[{\citenamefont{Wang et~al.}(2014)\citenamefont{Wang, Chen, Zhu, Wang,
  Zhu, Wu, Han, Zhang, Li, He et~al.}}]{QCAPA14}
\bibinfo{author}{\bibfnamefont{L.}~\bibnamefont{Wang}},
  \bibinfo{author}{\bibfnamefont{X.}~\bibnamefont{Chen}},
  \bibinfo{author}{\bibfnamefont{W.}~\bibnamefont{Zhu}},
  \bibinfo{author}{\bibfnamefont{Y.}~\bibnamefont{Wang}},
  \bibinfo{author}{\bibfnamefont{C.}~\bibnamefont{Zhu}},
  \bibinfo{author}{\bibfnamefont{Z.}~\bibnamefont{Wu}},
  \bibinfo{author}{\bibfnamefont{Y.}~\bibnamefont{Han}},
  \bibinfo{author}{\bibfnamefont{M.}~\bibnamefont{Zhang}},
  \bibinfo{author}{\bibfnamefont{W.}~\bibnamefont{Li}},
  \bibinfo{author}{\bibfnamefont{Y.}~\bibnamefont{He}}, \bibnamefont{et~al.},
  \bibinfo{journal}{Phys. Rev. B} \textbf{\bibinfo{volume}{89}},
  \bibinfo{pages}{075410} (\bibinfo{year}{2014}).

\bibitem[{\citenamefont{Bolotin et~al.}(2008)\citenamefont{Bolotin, Sikes,
  Hone, Stormer, and Kim}}]{Bolotin2008}
\bibinfo{author}{\bibfnamefont{K.~I.} \bibnamefont{Bolotin}},
  \bibinfo{author}{\bibfnamefont{K.~J.} \bibnamefont{Sikes}},
  \bibinfo{author}{\bibfnamefont{J.}~\bibnamefont{Hone}},
  \bibinfo{author}{\bibfnamefont{H.~L.} \bibnamefont{Stormer}},
  \bibnamefont{and} \bibinfo{author}{\bibfnamefont{P.}~\bibnamefont{Kim}},
  \bibinfo{journal}{Phys. Rev. Lett.} \textbf{\bibinfo{volume}{101}},
  \bibinfo{pages}{096802} (\bibinfo{year}{2008}).

\bibitem[{\citenamefont{Ren et~al.}(2012)\citenamefont{Ren, Zhang, Yao, Sun,
  Kaneko, Yan, Nanot, Jin, Kawayama, Tonouchi et~al.}}]{Ren2012terahertz}
\bibinfo{author}{\bibfnamefont{L.}~\bibnamefont{Ren}},
  \bibinfo{author}{\bibfnamefont{Q.}~\bibnamefont{Zhang}},
  \bibinfo{author}{\bibfnamefont{J.}~\bibnamefont{Yao}},
  \bibinfo{author}{\bibfnamefont{Z.}~\bibnamefont{Sun}},
  \bibinfo{author}{\bibfnamefont{R.}~\bibnamefont{Kaneko}},
  \bibinfo{author}{\bibfnamefont{Z.}~\bibnamefont{Yan}},
  \bibinfo{author}{\bibfnamefont{S.}~\bibnamefont{Nanot}},
  \bibinfo{author}{\bibfnamefont{Z.}~\bibnamefont{Jin}},
  \bibinfo{author}{\bibfnamefont{I.}~\bibnamefont{Kawayama}},
  \bibinfo{author}{\bibfnamefont{M.}~\bibnamefont{Tonouchi}},
  \bibnamefont{et~al.}, \bibinfo{journal}{Nano letters}
  \textbf{\bibinfo{volume}{12}}, \bibinfo{pages}{3711} (\bibinfo{year}{2012}).

\end{thebibliography}

\end{document}